\documentclass[a4paper,11pt]{article}
\usepackage[margin=1in]{geometry}
\usepackage{array}
\usepackage{caption}
\usepackage{subcaption}

\usepackage[english]{babel}
\usepackage[utf8]{inputenc}
\usepackage{amsmath}
\usepackage{amssymb}
\usepackage{mathrsfs}
\usepackage{eucal}
\usepackage{graphicx}
\usepackage{outlines}
\graphicspath{ {images/} }
\usepackage{fancyhdr}
\usepackage{lipsum}
\usepackage{listings}
\usepackage{xcolor}
\usepackage[]{algorithm2e}
\usepackage{nomencl}

\definecolor{codegreen}{rgb}{0,0.6,0}
\definecolor{codegray}{rgb}{0.5,0.5,0.5}
\definecolor{codepurple}{rgb}{0.58,0,0.82}
\definecolor{backcolour}{rgb}{0.95,0.95,0.92}

\lstdefinestyle{mystyle}{
    backgroundcolor=\color{backcolour},   
    commentstyle=\color{codegreen},
    keywordstyle=\color{magenta},
    numberstyle=\tiny\color{codegray},
    stringstyle=\color{codepurple},
    basicstyle=\ttfamily\footnotesize,
    breakatwhitespace=false,         
    breaklines=true,                 
    captionpos=b,                    
    keepspaces=true,                 
    numbers=left,                    
    numbersep=5pt,                  
    showspaces=false,                
    showstringspaces=false,
    showtabs=false,                  
    tabsize=2
}

\makeatletter
\newcommand{\thickhline}{%
    \noalign {\ifnum 0=`}\fi \hrule height 1pt
    \futurelet \reserved@a \@xhline
}
\newcolumntype{"}{@{\hskip\tabcolsep\vrule width 1pt\hskip\tabcolsep}}
\makeatother

\title{Deep Reinforcement Learning for Optimal Investment and Saving Strategy Selection in Heterogeneous Profiles: \\Intelligent Agents working towards retirement}

\author{
  Ozhamaratli, Fatih\\
  \texttt{ucabfoz@ucl.ac.uk}
  \and
  Barucca, Paolo\\
  \texttt{p.barucca@ucl.ac.uk}
}

\date{June 12, 2022}

\makenomenclature

\renewcommand{\nompreamble}{The next list describes several symbols that will be later used within the body of the document}

\begin{document}
\maketitle
\thispagestyle{empty}
\begin{abstract}
The transition from defined benefit to defined contribution pension plans shifts the responsibility for saving toward retirement from governments and institutions to the individuals. Determining optimal saving and investment strategy for individuals is paramount for stable financial stance and for avoiding poverty during work-life and retirement, and it is a particularly challenging task in a world where form of employment and income trajectory experienced by different occupation groups are highly diversified. We introduce a model in which agents learn optimal portfolio allocation and saving strategies that are suitable for their heterogeneous profiles. We use deep reinforcement learning to train agents. The environment is calibrated with occupation and age dependent income evolution dynamics. The research focuses on heterogeneous income trajectories dependent on agent profiles and incorporates the behavioural parameterisation of agents. The model provides a flexible methodology to estimate lifetime consumption and investment choices for heterogeneous profiles under varying scenarios.

\end{abstract}

\newpage
\tableofcontents
\thispagestyle{plain}
\setcounter{page}{1}
\newpage
\section{Introduction}
Retirement financing has been experiencing a clear transition trend from defined benefit (DB) schemes to defined contribution (DC) schemes, as reported by OECD \cite{oecd_pensions_2019},  \cite{oecd_pension_2019}, \cite{oecd_pension_2020}. DB schemes require scheme sponsors as ultimate guarantors of the schemes, and many DB schemes' deficit require their sponsors to bail out. Employers are preferring DC schemes, because the risk and responsibility of managing funds, longevity risk, and market risks are transferred to contributors in DC schemes. Furthermore, the contribution rates in DC schemes in the UK are on average significantly less, 5.1\%, in comparison to DB average contributions of 28.5\% \cite{ons_pension}. 
The effects of shocks during the accumulation phase are critical, some people were raiding retirement accounts amid COVID-19. Pensioners in under-pensioned groups \cite{wilkinson_what_2021} were draining pots and faced significant wage shocks, and this will have an effect on lower future cumulative pot and earnings, exceptional government policies were critical to alleviate the effects of COVID-19 on pension savings and wages but a significant shock with effects to the labour market could not be avoided.
It has become apparent how different professions can be affected differently by economic shocks, bringing the attention to the role of profile heterogeneity also in the context of pension management. 
For instance, the rise of the gig economy and irregular workforce participation modes enable more flexible work-life conditions, but introduce larger variations to income trajectories due to the lack of guaranteed income streams. 

Previous research has addressed the income distribution and its relationship with age \cite{ozhamaratli_generative_2022}, that can be used to investigate the effects of demographic shifts and aging population on income.
The increasing heterogeneity of career paths and income trajectories requires addressing the questions of how much to save in a more granular way, as well as how to allocate the savings between spendable liquid investments and illiquid retirement investments. 
In the literature the life-cycle models of income, consumption, and portfolio allocation have been analysed with various perspectives.
Samuelson approached lifetime portfolio selection \cite{samuelson_lifetime_1969} as dynamic stochastic programming in discrete time and solved the many-period generalisation corresponding to lifetime planning of consumption and investment decisions. 
Merton formulated the continuous-time version \cite{merton_lifetime_1969} of the same approach for portfolio selection under uncertainty. Later he extended these results \cite{merton_optimum_1971} to more general utility functions, price behaviour assumptions, and for income generated also from non- capital gains sources.
A comprehensive study \cite{cocco_consumption_2005} proposes a life-cycle model of consumption and portfolio choice, as an intertemporal portfolio optimisation problem where labor income is assumed to be a risk-free asset, and calibrates it with real-world data by solving numerically, where asset choices are risky and risk-free.
Coco et al. \cite{cocco_consumption_2005} present a model where risky income is invested in either risky asset or riskless asset, both are liquid and can be used for consumption, they model the income process explicitly and analytically; they solve the optimal portfolio allocation problem at a given age by numerical solution of their model with backward induction. A following study by Campanale et. al. \cite{campanale_life-cycle_2015} present a model, which includes explicit formulation of income process; but differs from previous research by introducing liquidity friction to risky asset, by charging an excess cost if consumption is financed through the risky asset. The model must be solved numerically via special algorithm, and the solution is described by authors as slow and difficulty, due to three continuous state variables, two continuous controls, and fixed transaction cost breaking the concavity of the objective function. The Campanale model assumes that a person has the freedom to switch between liquid and illiquid asset types, which is not the case with locked pension savings. Campanale et al. uses dynamic programming to optimise the Epstein-Zinn\cite{epstein_substitution_1989} preference utility of a household, given specific labour income process consisting of deterministic G(t) of third-order polynomial, and idiosyncratic shock.
In the Campanale et al. model, the most important calibration challenge is the transaction cost, which also includes psychological and nonmonetary costs. 

Further studies focus on liquid and non-liquid retirement savings accounts where liquidity is constrained by introducing cost to liquidate retirement savings\cite{dahlquist_asset_2016} and \cite{campanale_life-cycle_2015}.
Previous research fails to address the heterogeneity of contributor profiles and falls short of addressing the idiosyncratic challenges of avoiding consumption crisis during unemployment periods and saving an adequate pension pot for retirement.

Advances in agent-based modelling of complex financial systems, increased computational power, and advances in techniques for optimising agent behaviour in complex environments are enablers and drivers of the research. Our research uses a novel methodology to address complex problems of financial systems, one such effort in the recent literature is the model called AI Economist\cite{zheng_ai_2020}, which uses AI-assisted deep reinforcement learning and implements an agent-based model to address the needs of socioeconomic challenges introduced by designing and testing economic policies, where modules called social planners are trained to discover tax policies in dynamic economies that can effectively trade off economic equality and productivity. A two-level deep reinforcement learning approach is applied to learn dynamic tax policies, based on economic simulations in which both agents and a government learn and adapt.

We introduce a simple model of contributor agents, who decide how much to save and how to allocate the savings, this decision is effected by simplified behavioural dynamics and information flow in the peer network. Agents decide and optimise their allocation strategy using a deep neural network trained with reinforcement learning. We introduce a simple simulation environment for the agents; which encapsulates employment and income dynamics. Our research represents a bridge between agent-based modelling of the pension system and deep reinforcement learning for finance.

We provide results from trained agents reflecting behaviour in par with the literature on aggregate level and agent-specific optimal behaviour with higher granularity for heterogeneous profiles. The model is dynamic, scalable, and can be calibrated to different scenarios. The results show that the balance between near-term consumption safety and retirement savings can be achieved by profile-specific allocation strategies.

Our research differs from previous studies also in terms of extensive behavioural modeling and parameterisation of the agents. It captures effect of information transmission and emphasizes consumption sensitivity against negative shocks, and covers the utility perception.

In addition, our model makes it possible for contributors to account for occupation-specific dynamics of life-time income trajectories, which in turn makes it possible to prepare against profile-specific income shocks by allocating savings to cash buffer at the right time frames of their lives.

Our research represents a significant first step to model pension finances in an agent-based model with deep reinforcement learning which permits model configurations with increased complexity and realism, in our paper we presented a simple two asset version with simple environment dynamics. 

\section{Model}

We introduce a simple model where the agents interact with the simulation environment and optimise their savings behaviour. During each cycle, agents observe the environment in which they are situated; they choose to allocate their income between consumption, liquid and illiquid assets. 
 
Each agent has a heterogeneous profile reflecting the occupation and demographic characteristics; these characteristics are determinants of the unique income and consumption trajectories of each agent. Agents also have characteristic behavioural parameters such as shock sensitivity, consumption utility, and peer-influence factor; which effect the way agents perceive the world and assign value to their stances. In particular, agents are bootstrapped in a social graph which is used for the transmission of information, such as employment status. 

Each month, agents receive their income according to their employment. Simulated employment and market dynamics, such as asset return rates, are exogenous and provided by the modeler according to empirical observations. The employment dynamics is dependent on heterogeneous profiles (occupation and demography) and includes the new employment of unemployed agents.

The agent first decides how much to save and how much to consume, and secondly the agent allocates the saved amount among a liquid asset and illiquid asset towards pension savings, each with different return rates. In order to make this financial decision, the agent's profile, income, behavioural parameters, and peer information observed from the own social network are given as an input to a deep policy network.

Deep reinforcement learning and parallel simulation of nearly 30000 agents in 100M timesteps is used for training the deep policy network. The policy network learns an optimal saving and investment strategy for pension savings and avoiding a consumption crisis during unemployment due to insufficient liquid savings.

\subsection{Agent and Environment Cycle}
In order for the simulation to be integrated with existing frameworks, the AEC(Agent Environment Cycle)\cite{terry2020pettingzoo} is followed to also provide a standardised GYM-Like API. The simulations are vectorised and run in parallel. For the purpose of this research, the simulations are conducted in parallel utilising 32 processors, where each processor runs a cohort of more than a thousand agents. For each time step, all of the agents observe and act simultaneously. 

Agents observe the environment; these observations include information regarding the market, graph, and agent's own state including occupation, age, income, and wealth.

The agent action $a_i,t$ is shaped by policy $\pi_i$, during learning the reward $r_i,t$ for the agent is the sum of total discounted utility and penalty for consumption crisis, which denotes the situation where the agent cannot finance its consumption $c_{i,t}$ governed by consumption dynamic $C$. Agent behaviour is shaped by influences from peers, individuality factor, consumption utility, and shock response characteristics.

The agent policies are modelled with a deep neural network, which takes as input agent specific observations and hidden-state:
\begin{equation}
a_{i,t} \sim \pi(o_{i,t}^{network},o_{i,t}^{agent},o_{i,t}^{market},h_{i,t};\theta)
\end{equation}
The parameter variable $\theta$ is not agent specific, but common for all contributor agents and the hidden state is updated during action inference of policy network.

\begin{itemize}
\item $o_{i,t}^{network}$: Observation of the network
\item $o_{i,t}^{agent}$   : Observation of own behavioural factors, income, and resources.
\item $o_{i,t}^{market}$ : Observation of the market, such as 
\item $h_i,t$      : Hidden state. The updating of hidden state can be interpreted as agents updating their Risk Profile given observations and previous state. And in the future the hidden-state can be used as Risk Profile embedding.
\end{itemize}

The action space is as follows: 
\begin{itemize}
\item $a_{i,t}^{save}$      : Decides to save $x\%$ (and consuming $(100-x)\%$) 
\item $a_{i,t}^{liquid}$    : Decide to allocate $y\%$ to liquid asset $x\%$ (and allocating the illiquid asset $(100-x)\%$) .
\item Saving and liquidity percentages are discretised into bins such as $[0, 0.25, 0.5, 0.75, 1]$ in the model.
\end{itemize}

The full list of variables can be found in the Appendix.
\begin{figure}[htp]%
         \centering
         \includegraphics[width=0.73\linewidth]{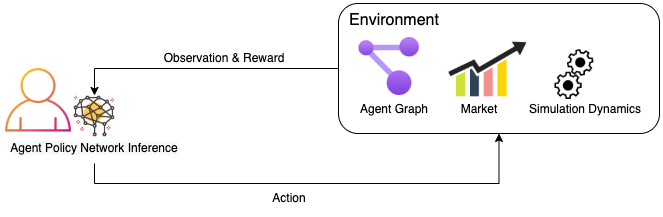}
         \caption{Agent and Environment}
         \label{fig:model_stack}
\end{figure}

The utility of an agent is expressed as the sum of the utility of consumption and of savings. 
\begin{equation}
u_{i,t}=u_{i,t}^{consumption}+u_{i,t}^{savings}
\end{equation}
Both the consumption and the savings utility functions are considered linear in this study. 
 Constant relative risk aversion(CRRA) \cite{pratt_risk_1964} is defined as : 
\begin{equation}
crra(l,\eta)= \begin{cases}\frac{l^{1-\eta}-1}{1-\eta} & \eta \geq 0, \eta \neq 1 \\ \ln (l) & \eta=1\end{cases}
\end{equation}
Individual consumption is calculated by multiplying the consumed amount with agent specific consumption utility factor,

\begin{equation}
u_{i,t}^{consumption} = \sum_{t=0\dots t} \left[crra(\mathbb{C}_{i,t},q_{i,t})\right]
\end{equation}

The utility of the saved amount equals the monetary value of the total savings.
\begin{equation}
u_{i,t}^{savings} = x_{i,t}
\end{equation}
If the agent is unemployed or allocated insufficient funds to fulfil minimum consumption required by the modeller, then the liquid funds are used to finance consumption. If the funds are insufficient, a consumption crisis occurs, which impacts reward negatively with a consumption crisis penalty. 
If the agent consumes a lesser percentage then it is required to finance at least minimum consumption limit, then the invalid action penalty occurs.
Consumption crisis is penalised if liquid savings cannot finance the decided consumption at time step t:
\[
\mathbb{P}_{i,t} = 
\begin{cases}
\psi & \text{if } \mathbb{C}_{i,t}>x^{liquid}_{i,t} \\ 
0 & \text{if } \mathbb{C}_{i,t}\leq x^{liquid}_{i,t} 
\end{cases}
\]
Invalid Actions, such as insufficient consumption rate allocation, are penalised if the decided consumption is less than minimum consumption limit despite having excess earnings that could be used for financing the minimum consumption amount:
\[
\mathbb{AP}_{i,t} = 
\begin{cases}
(m-\mathbb{C}_{i,t}) * \zeta & \text{if } m>\mathbb{C}_{i,t} \ and \  \mathbb{E}_{i,t}>\mathbb{C}_{i,t} \\ 
0 & \text{else } 
\end{cases}
\]
The results of the actions are evaluated as utility difference of the new state of the agents, which constitutes the raw rewards of the actions.
\begin{equation}
\Delta_{i,t} = u_{i,t}-u_{i,t-1}
\end{equation}
The consumption crisis and invalid action penalties are used to augment the utility difference to constitute the reward.

\begin{equation}
r_{i,t} = \Delta_{i,t} * f(\Delta_{i,t},\kappa_{i,t})-\mathbb{P}_{i,t}-\mathbb{AP}_{i,t}
\end{equation}The negative utility difference is augmented with an agent's shock perception modifier, in order to amplify the negative shocks according to the agent's behavioural parameter.
\[
f(\Delta,\kappa) = 
\begin{cases}
1 & \text{if } \Delta\geq 0 \\ 
e^{\kappa} & \text{if } \Delta< 0 
\end{cases}
\]
\begin{equation}
r_{i,0} = u_{i,0}^{savings}
\end{equation}

The actions are percentage choices between consumption and savings, and investment choices between pension orientated non-liquid funds and liquid funds that can be used at any time to finance consumption, these funds have a vital function especially during the times of unemployment.

The AI Economist's approach\cite{zheng_ai_2020} stated as a multi-agent Markov Game can be defined by the tuple ($S$, $A$, $r$, $\mathscr{T}$, $\gamma$, $o$, $\mathcal{I}$), for the State Space, Action Space, Reward, State Transition Distribution, Utility Discount Rate, observation, and Agent Indices.
The agents observe $o_{i,t}$, which is by the agent $i$ observable fraction of the state $s_{i,t}$. The agent learns a policy, $\pi(a_{i,t}|o_{i,t},h_{i,t};\theta_{i})$, and executes an action $a_{i,t}$ according to the policy. The state of the environment is updated according to $\mathscr{T}(s_{t+1}|s_t,a_t)$. Agents maximise their $\gamma$ discounted expected return, depending on the hidden state $h_{i,t}$, and the policy parameter $\theta_{i}$. Following the same reinforcement learning approach as AI Economist, the agent maximises its expected reward, depending on the actions $a_{-i}$ and behavioural policies $\pi_{-i}$ of the other agents, and environment transitions:
\begin{equation}
\max\limits_{\theta_{i}}  \boldsymbol{E}_{a_i\sim\pi_i,a_{-i}\sim\pi_{-i},s'\sim\mathscr{T}} \left[\sum_{t}\gamma^{t}r_{i,t} \right]
\label{eq:policy_optimisation}
\end{equation}
During action execution, the rewards and utilities are calculated and the tuples of observation, hidden state, and action vectors, as well as rewards are recorded to replay buffer, for future training. The actions are sometimes random for greedy exploration of the domain, the epsilon-greedy parameter is provided by the modeller. At regular intervals, the replay buffer is used for training the deep policy network by utilising the rewards with policy gradients.
The model is trained with A2C, PPO and Policy Gradient methods. A2C training results in slightly better results, which are presented in the article.

\subsection{Behavioural Parameters of Agents}
For modeling behaviour, we base our parameterisation on the approach in \cite{cane_validation_2012}, where the authors investigated the applicability of the Theoretical Domains Framework outside clinical uses for cross-disciplinary implementation and other research on behaviour change, and provided a simplified version containing 14 domains and 84 component constructs. Theoretical Domains Framework is wide and reports on pension behaviour tend to focus on few factors; for the scope of our research we chose three factors:

\begin{itemize}
\item Consumption Utility: How do they value current consumption? An agent specific consumption utility multiplier factor
\item Shock Response Characteristics: How do they respond to the shock? A factor reflecting how sharp do agents react to the shock and how drastic are they decreasing their consumption.
\item Individuality Factor: How are they being affected by each others beliefs and decisions.
\end{itemize}

Some agents are optimistic and underestimate the severity of the shocks, and some agents are pessimist and overestimate the effect of the shocks. The shock sensitivity factor $\kappa_{i,t}$ is a multiplier of the perceived shock effect, which is normalised for agents of the same occupation. It can be assigned from a normal distribution, can be controlled for experimentation, or fed from empirical report.
The agents are affected by the peers and the shocks experienced by the peers(if $z_{i,t-1}=0$). The observations are informative for the closer peers on the graph, and becomes less informative for other agents with weaker connection on the graph. The shocks that effect the peers are also weighted with the shock-sensitiveness parameter.
The behaviour parameters that are introduced in this section are fixed during the entire simulation, but later research can adapt the model to allow the parameters to change during lifetime.

\subsection{Deep Policy Network for Optimal Saving, Investment and Liquidity}

Agent observations are expressed as a single vector that comprises the concatenation of agent, market and graph vectors. The observation vector is passed through the deep neural network towards the LSTM, which updates the agent's hidden state and outputs a vector for next layer, which is softmaxed to output a vector representing the action probabilities. A single policy network is trained for all actions: The action can be as follows "('C25', 'L75')", where "C25" means consume 25\% and save 75\%; "L75" means allocate 75\% of your savings to liquid asset, and 25\% of your saving to nonliquid asset. 
 
 \begin{figure}
  \centerline{\includegraphics[scale=0.68]{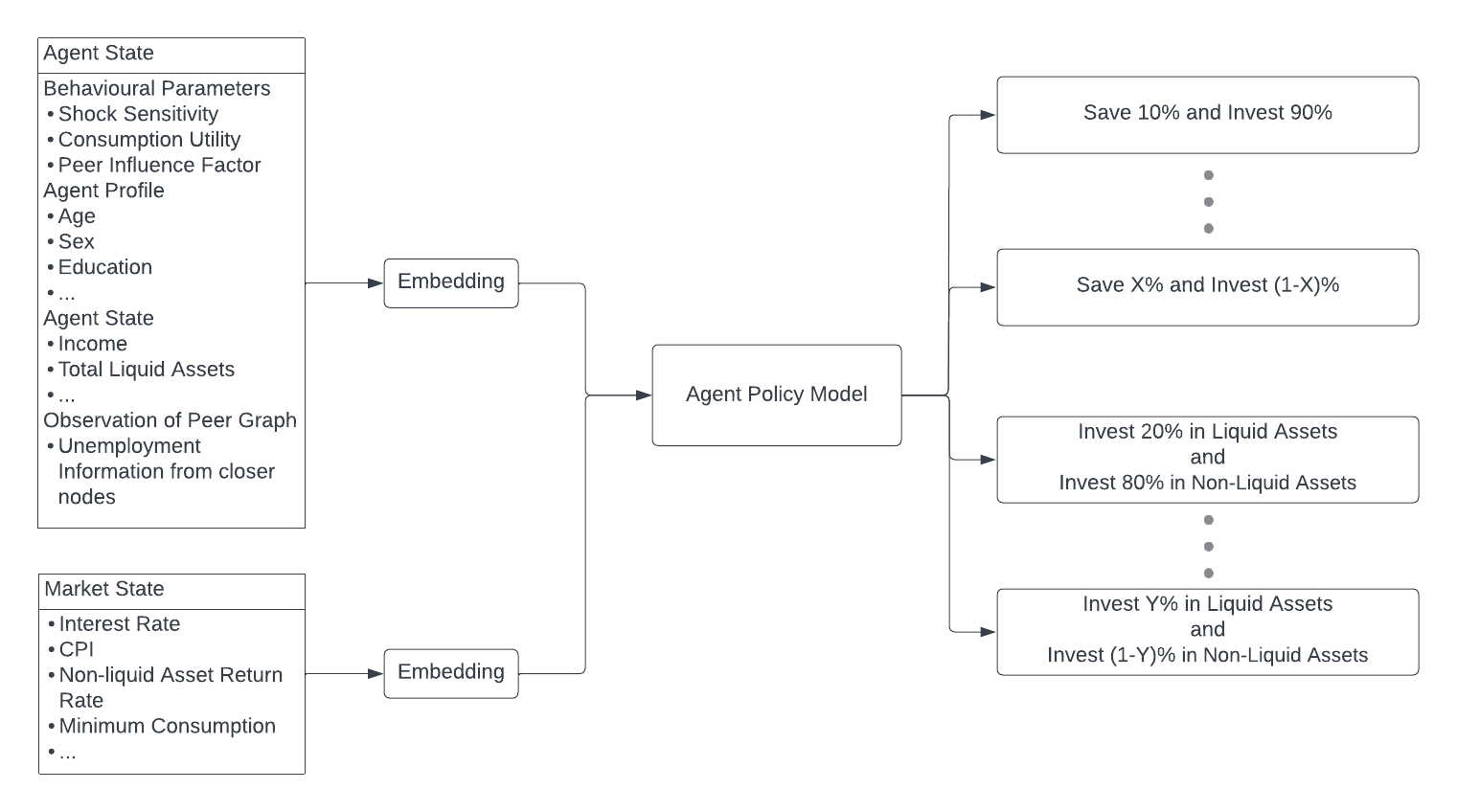}}
  \caption{Policy Model}
  \label{fig:policy_model}
\end{figure}

The hidden states from the model can be thought of as Risk Profile Embedding, which is updated by observations and processing the Agent Profile with the observed environment and shocks via a Deep Neural Architecture that can be found on Fig. \ref{fig:policy_model}. Reinforcement Learning is used for adjusting the allocation profile according to the Risk Profile Embedding also expressed as a hidden state.
At each time step, the agent decides to allocate the income among consumption, savings, and investment classes. This is accomplished by a deep neural network constituted of several layers of a Feedforward Neural Network with nonlinearity as well as a recurrent unit of LSTM, which is responsible for acting as the memory of agents, and the hidden states can be used as agent representations during further studies. The details of the neural architecture can be found in the Appendix. 

There is a single action space unifying the choices of consumption and liquidity preference, which means that there are not two different networks for different decisions but one unified network which represents the collection of actions such as "('C25', 'L75')".
An alternative choice was continuous action space of two different actions expressed as a vector of length two, which is expected to have more capability to capture relationship between similar amounts, especially if the vast amounts of simulation data are not available. Furthermore, the dimension of the action space is greatly reduced if the number of discrete consumption and investment type rate choices are high.
Setting the reward function for the agents is the trickiest part of the training process, different reward function structures tend to give spurious and unintended conclusions especially due to the existence of both positive-effecting factors and negative-effecting factors, which makes the hyper-parameter tuning for the penalties paramount; and requires several trials to find the hyper-parameter enabling successful optimisation of the agents. Failing to tune the penalties results in unintended shortcuts that obstruct the main goal of optimising agent behaviour in an understandable and meaningful way.

\section{The Environment}

At each time step, first the environment operations are executed.
Agent environment operations are executed as follows: first, the market dynamics are executed which ensures that assets are gaining value according to the calculated interest rates determined by the modeler.
Secondly, essential population dynamics is executed such as ageing of agents, executing death operations according to the age-specific death probability.
The Retirement Process checks if any new agents are required to retire due to age. If an agent retires, their retirement pension is calculated as a rate of their previous consumption at employment, according to the recommended guidelines of the OECD.
If an agent is retired, then the agent collects pension from illiquid pension fund that they contributed during employment life.
The agents that are not retired are processed to determine stochastically if they will lose their employment and, if so, for how long they will stay unemployed according to the unemployment duration distribution dependent on the occupation and age.
Unemployed agents are assigned new income at their new jobs according to the income distribution depending on occupation and age. These distributions are fed as quantiled distribution tables to simulation.
The employed agents receive their salary at each month according to their predetermined income.

The agents decide how to allocate their income between consuming and saving, and decide to allocate the saved amount in liquid and riskless assets, or illiquid and low-risk assets. The decision is shaped by learnt policy, and observations, which includes the market dynamics, information regarding actions and behaviour of peers, and considering the agent's own profile.

\subsection{The Graph and Synthetic Population}
A synthetic representative population is used for the initialising agent population, information such as age, income, profession, education level, and other relevant background information are included.

We assume the employee network consists of three communities divided by income level as low, medium, high, and the three communities have significant intercommunity interaction but limited intracommunity interaction. The graph choice is based on the idea that geographical and social networks are also characterised by the socioeconomic clusters, and the choice of three communities with income levels is the simplification of the socioeconomic network. The synthetic database is generated according to the basic insights from the surveys. Later investigation could incorporate survey data to bootstrap the population and investigate geographical graph, potential social network data, and known network structures to model connection between agents.

Observation of graphs can be done in several ways; a simplistic way to do is modelling information transmission between each agent and its vicinity; for the first iteration of the model this means that the neighbours and their neighbours, and transmission of employment information. A more advanced graph observation might be modelled as transmission of not just employment information, but also incorporating additional information such as occupation and the income or consumption data; moreover, the near-neighbour graph can be represented with state-of-the-art graph embedding methodology.
\begin{equation}
o_{a,t}^{network} = \sum_{a,b}\left[\sum_{b,c} A_{b,c}\delta(\mathbb{E}_{c,t},0)\right] + \sum_{a,b} A_{a,b}\delta(\mathbb{E}_{b,t},0)
\end{equation}
For the purpose of experimentation and investigation of the model, a synthetic but representative population can provide both fidelity and flexibility in a controlled environment. As a design choice for the synthetic population network, we include three clusters, which can be thought of as three neighbourhoods; these neighbourhoods possess nodes with three different income groups: high, medium, and low income. 
Each node is connected to its own neighbourhood node, and the neighbourhoods are connected to each other with specified weights. 
Agents are bootstrapped with one of the general occupation groups, occupation-specific incomes, employment status, and ages derived from USA Census Data\cite{bls_2019}. Census data are used to generate the synthetic agent population.

\subsection{Simulation Processes}
The simulation is initialised by bootstrapping the agent population and processes. During each time step, the simulation dynamics such as getting income, getting employed if vacant, are applied first, then the agent decides to allocate income for the consumption or saving, and decides to save by investing in liquid assets, which can be liquidised easily during unemployment, or illiquid assets which are towards a future retirement, but usually have better return. Agents are bound by constraints such as the need to consume a minimum amount determined in light of government statistics\cite{aspe_2019} that determines a minimum consumption per individual.

The occupation-specific income for new employment is determined according to the summary tables from USA Census Data. The tables reflect the quantile breakdown, and the agents are probabilistically assigned to one of the income quantiles.

The unemployment events and employment processes are explicitly modelled and calibrated with the US Census Data\cite{bls_2019}. The probability of unemployment and the duration of unemployment are determined according to the summary tables of the US Census.

Retirement age and retirement income can be accounted for in the system. For the sake of simplicity, initial simulations neglect the retirement period, by only focusing contribution period; but the system is later extended to cover the retirement period. Retirement income is defined as a fraction of the last income, fractional retirement income is recommended by international institutions, and this methodology is often also used in the literature\cite{cocco_consumption_2005}.

The agent death probabilities are modelled using the Actuarial Life Table \cite{ssa_2017} in order to make the model comparable with existing models in the literature.

\subsection{Scaling}
The agent observations are continuously scaled and standardised, with an online methodology. This is due to the fact that the training dataset is generated continuously during simulation and the distribution of the observed dataset is not known in advance at the start of the simulation, but it can be learnt to an extent after several epochs, and these learnt scales can be utilised in following training and inference as well.
The relevant agent variables("OCC\_CODE", "income", "consumption\_utility\_factor", "shock\_sensitivity\_factor", "individuality\_factor", "non\_liquid\_asset", "liquid\_asset") are transformed to a vector by concatenating categorical one hot vectors with the values of the continuous variables; here the standardisation of these categorical variables is challenging due to the variability of the aspects such as accumulated liquid assets. The huge value differences have the potential to introduce instability during the training of machine learning models.
The market state captures important variables such as interest rates given to different asset classes as a dictionary, the market dictionary is transformed to a vector as well. 
The perception of graph by the agent is represented as a vector, for the sake of simplicity in this first version of the model the graph is represented by single-valued vector, which is calculated by second-order graph traversal of neighbouring agents in the network, the income information from neighbouring agents is incorporated according to the edge weights, as well as the individuality factor of the effected agent.

\section{Results}
We look at longitudinal trajectory plots and strategy breakdown per total asset size, which provide granular information regarding the differences between occupations. 
These plots can capture various scenarios such as differences between early career and mid career saving rate strategies among various occupations; which provide more tailored strategies for short-term consumption security and healthy long-term pension finances.
22 parallel initially identical cohorts are simulated for 1000 weeks of agent-time in order to generate resulting tables and plots, which results in 40M agent time-step samples.

\subsection{Labour, Income, Consumption and Wealth}

\begin{figure}[htp]%
         \centering
      \includegraphics[width=1\linewidth]{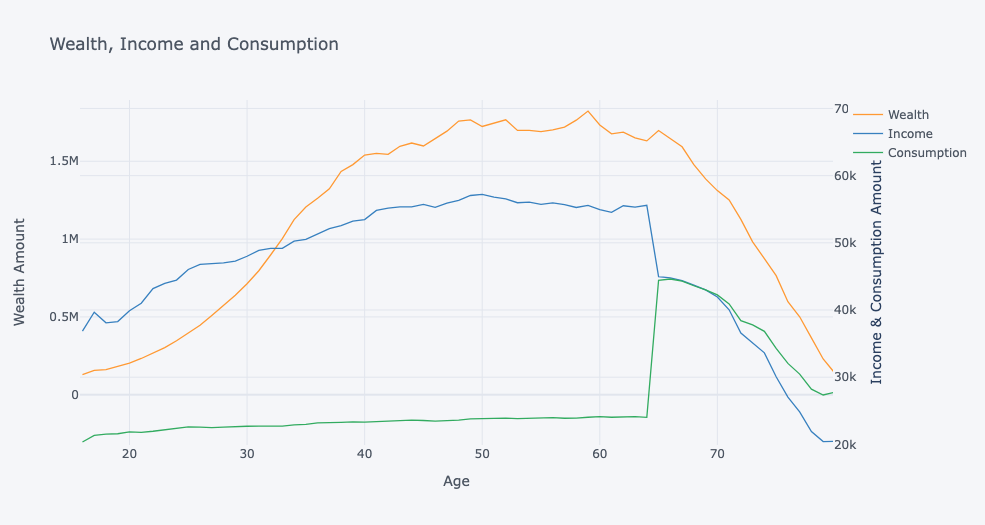}
         \caption{Wealth, consumption and labour income vs age plot}
         \label{fig:wealth_income_consumption_frac}
\end{figure}

 Fig. \ref{fig:wealth_income_consumption_frac} reflects a similar shape of average simulated income, consumption, and wealth accumulation and decrease over the life cycle compared to Coco et al. \cite{cocco_consumption_2005}. 
 The simulated income trajectory is a reflection of the observed data, which is used for calibration of the environment, and the shape of decrease by retirement age is due to the retirement income being defined as a fraction of last income, which then gradually decreases. The consumption trajectory during the work-life reflects saving choices of the population. The agent saves during work-life for financing potential unemployment periods and for retirement finances. The pension income and consumption at retirement age of 65 converges to the determined retirement income percentage of 80\% of latest salary. The data becomes noisy for older ages of 80, which might be due to significantly smaller sample size.
 
During retirement the pension income is supposed to come from pension savings that have been non-liquid during work-life, but if the pension savings are depleted any liquid savings can be used to finance the retirement income on Fig. \ref{fig:liquid_nonliquid_assets_by_ocupation_at_age}.
An interesting outcome of mandating pension income at retirement to be 80\% of employment income is comparatively lower consumption during employment, which might not be desirable; but our optimiser were forced to high saving rates due to 80\% mandate, which is stipulated by literature, detailed information can be found in previous sections focusing on literature. There might be various solutions to this problem that are out of our scope, such as easing pension level mandate, or government contributions, or higher returns of investment; these potential solutions can be investigated in future research.
 
 \begin{figure}[htp]%
         \centering
         \subfloat[Liquid Asset]{\includegraphics[width=0.5\linewidth]{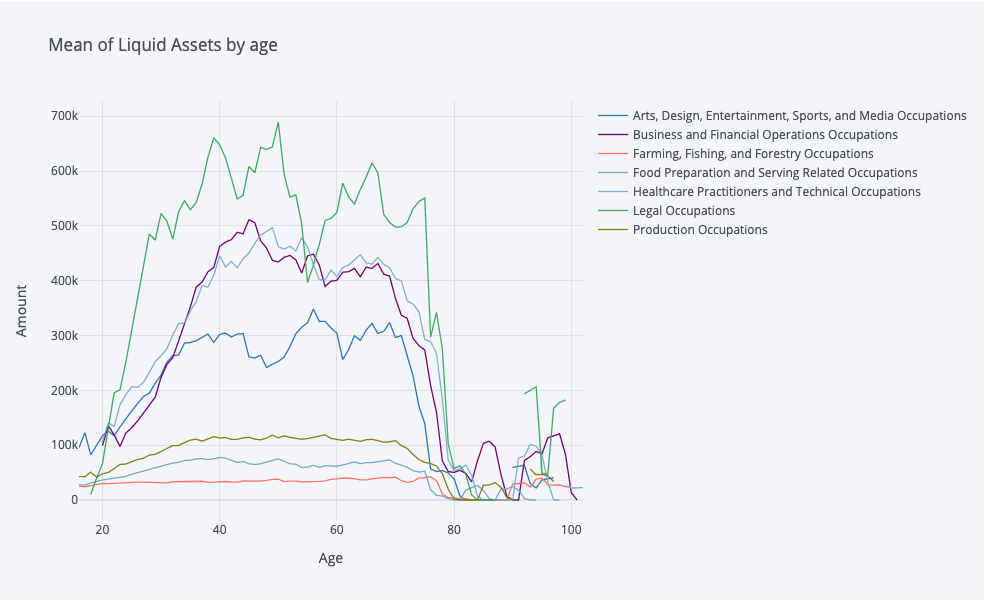}}
         \subfloat[Non-Liquid Asset]{\includegraphics[width=0.5\linewidth]{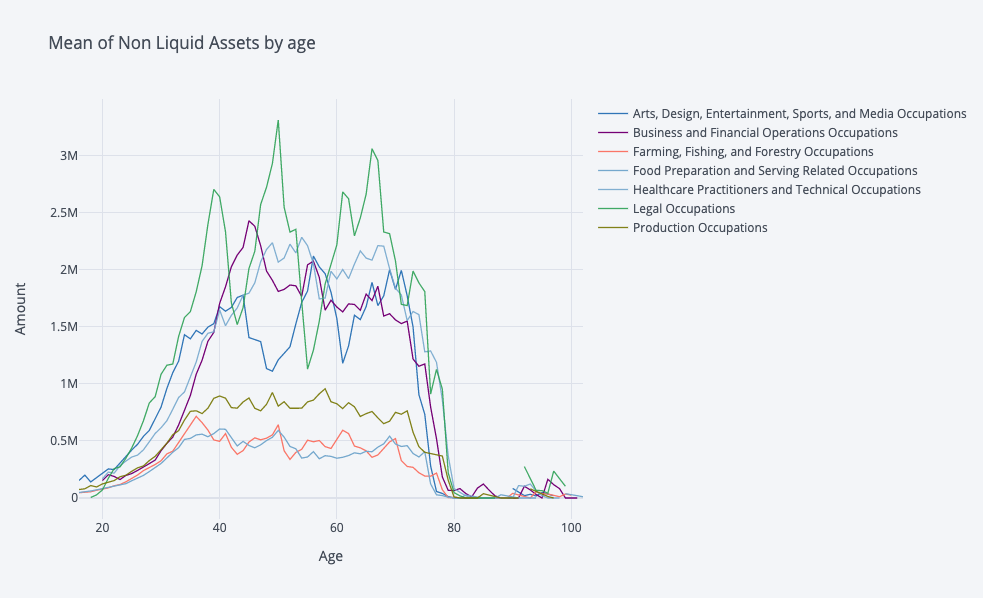}}
         \caption{Liquid and Nonliquid asset amounts by occupation at age, where only a selection of occupations are depicted on plot for clear visibility. The different characteristics of occupation groups are reflected by plots. }
         \label{fig:liquid_nonliquid_assets_by_ocupation_at_age}
\end{figure}

\subsection{Saving Profiles}
\begin{figure}[htp]%
         \centering
         \subfloat[mean income]{\includegraphics[width=0.5\linewidth]{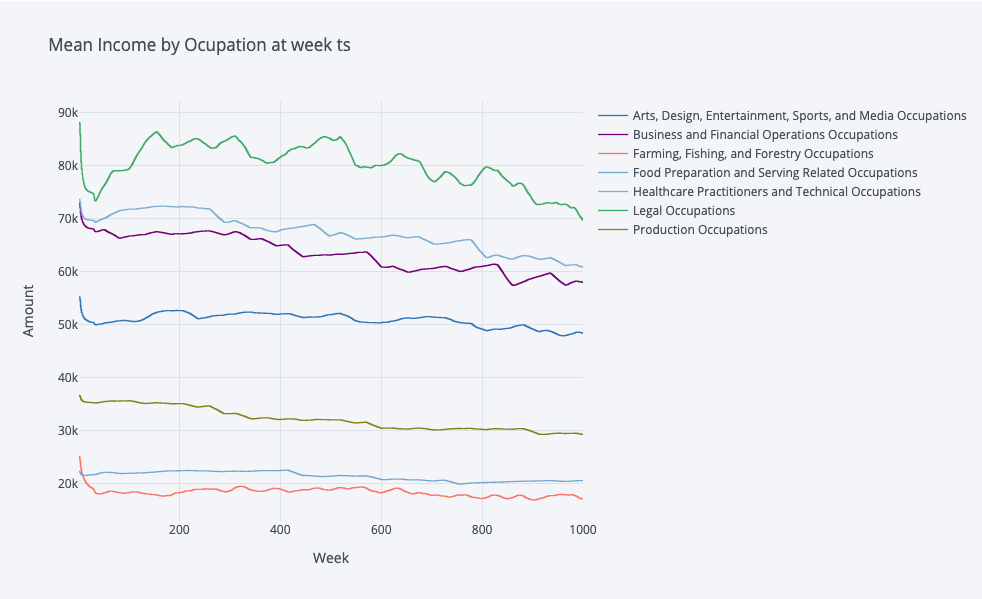}}
         \subfloat[unemployment]{\includegraphics[width=0.5\linewidth]{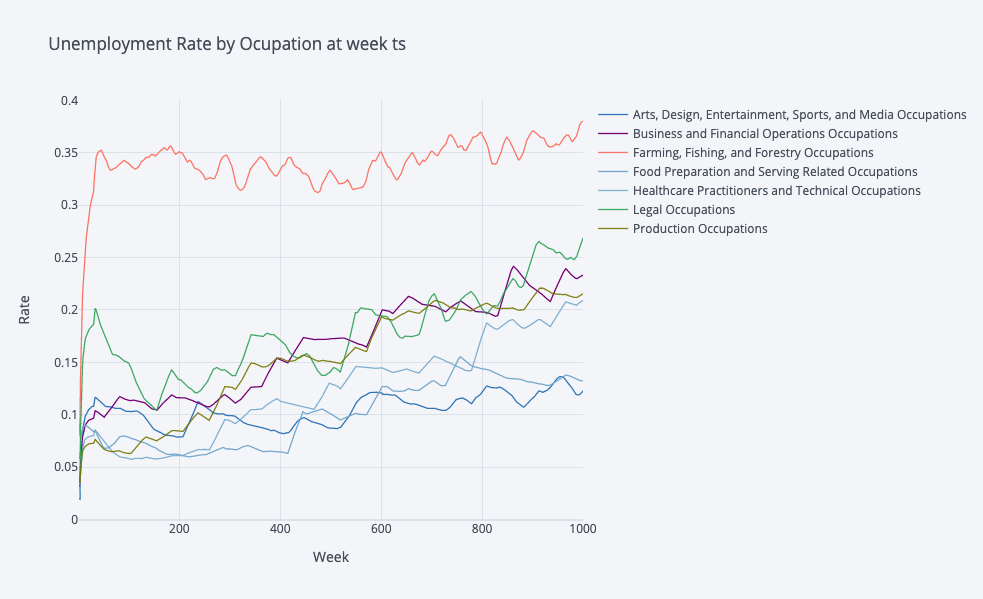}}
         \caption{Mean of income and unemployment by occupation at week ts; the values are smoothed by 30-step moving average and only a selection of occupations are depicted on plot for clear visibility. The different characteristics of occupation groups are reflected by plots.}
         \label{fig:income_unemployment_by_occ_at_week}
\end{figure}

The evolution of occupational income in a time frame of nearly 20 years on Fig. \ref{fig:income_unemployment_by_occ_at_week} reflect different characteristics for each occupational group, occupations such as "Sales and Related" and "Transportation and Material Moving" reflect significantly lower mean income with lower variance characteristics; on the contrary, occupations such as "Legal" and "Management", which reflect highest mean income and high variance of income for each occupation group. This plot reflects even at the simplest level that the income characteristics of each occupation can differ greatly.
The unemployment characteristics reflect great diversity, where occupations such as "Farming, Fishing and Forestry" possess greater and fluctuating risk profiles, which might be partially due to the characteristics of seasonality in these specific occupations.
No obvious dependence of saving rate or non-liquid investment rate on age or income level can be found in the analysis, showing the complexity of the decision making happening in the system. 

\begin{figure}[htp!]%
         \centering
         \subfloat[Week]{\includegraphics[width=0.5\linewidth]{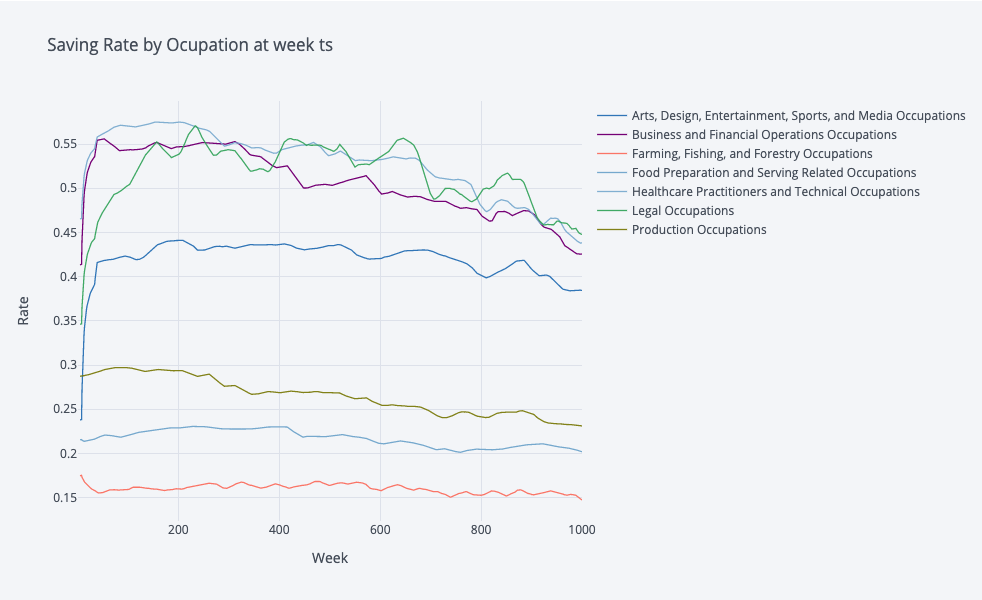}}
         \subfloat[Total Asset Amount]{\includegraphics[width=0.5\linewidth]{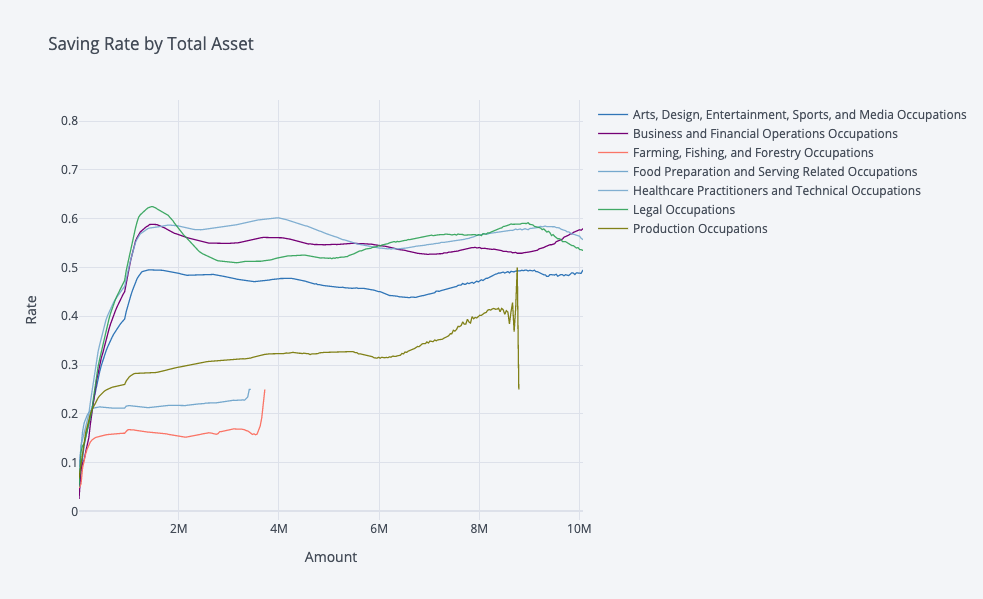}}
         \caption{Saving rate by occupation at week ts and saving rate by occupation at amount capped at 10M, the values are smoothed by 30-step moving average and only a selection of occupations are depicted on plot for clear visibility.}
         \label{fig:saving_rate_by_occ_at_week_asset}
\end{figure}

The savings profiles in Fig. \ref{fig:saving_rate_by_occ_at_week_asset} reflect heterogeneous characteristics, where at the same total wealth the saving rate differs greatly, which can be due to different income levels and unemployment risks of occupations. The saving rate plot shows increasing noise at higher wealth levels near 10M, and a much clearer trajectory at lower wealth. An interesting insight is that at the lowest wealth levels, all occupations display similar saving rates.
Minimum consumption requirement has a direct consequence of lower saving rates by occupations with low income
occupations such as "Farming, Fishing and Forestry", "Building and Grounds Cleaning and Maintenance", "Personal Care and Services", "Food Preparation and Serving Related" occupations have very low saving rate due to their difficulties to finance minimum consumption. 
Some general patterns can be identified, such as lower income occupations tend to have lower saving rates, but it does not imply that income itself can explain saving decisions; as we can observe varying saving rates among "Healthcare Practitioners", "Legal Professionals", and "Business and Financial Operations". 

\subsection{Portfolio Allocation}
\begin{figure}[htp]%
         \centering
         \subfloat[Total Asset Amount]{\includegraphics[width=0.5\linewidth]{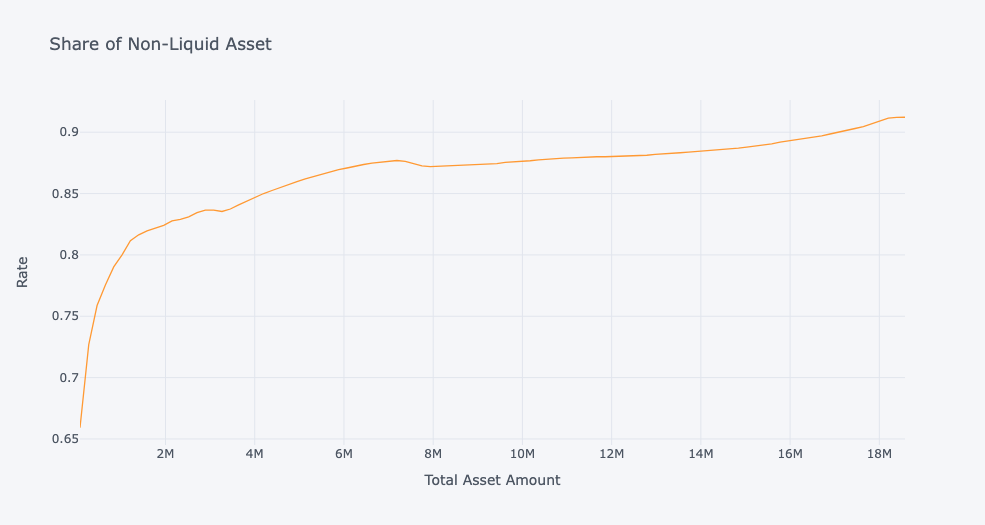}}
         \subfloat[Age]{\includegraphics[width=0.5\linewidth]{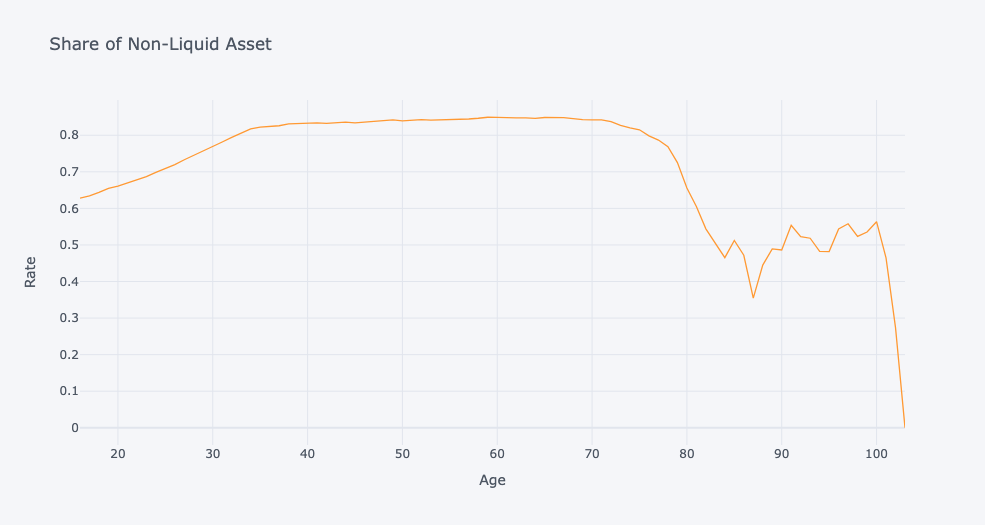}}
         \caption{Nonliquid asset share vs total asset amount and age}
         \label{fig:nonliquid_asset_share_vs_total_asset_amount_age}
\end{figure}

Compared to the literature on the share of non-liquid assets and age distribution, Fig. \ref{fig:nonliquid_asset_share_vs_total_asset_amount_age} reflects similar shape and rates, where Campanale et al.  \cite{campanale_life-cycle_2015} plots a very similar life-cycle stock share model profiles when the illiquid and liquid assets are differentiated with transaction costs for switching between them, our model reflect similar characteristics when the transaction costs are high. The share of non-liquid asset according to total current wealth also reflects a similar shape of initial increase followed by plateau. 

\begin{figure}[htp]%
         \centering
      \includegraphics[width=0.55\linewidth]{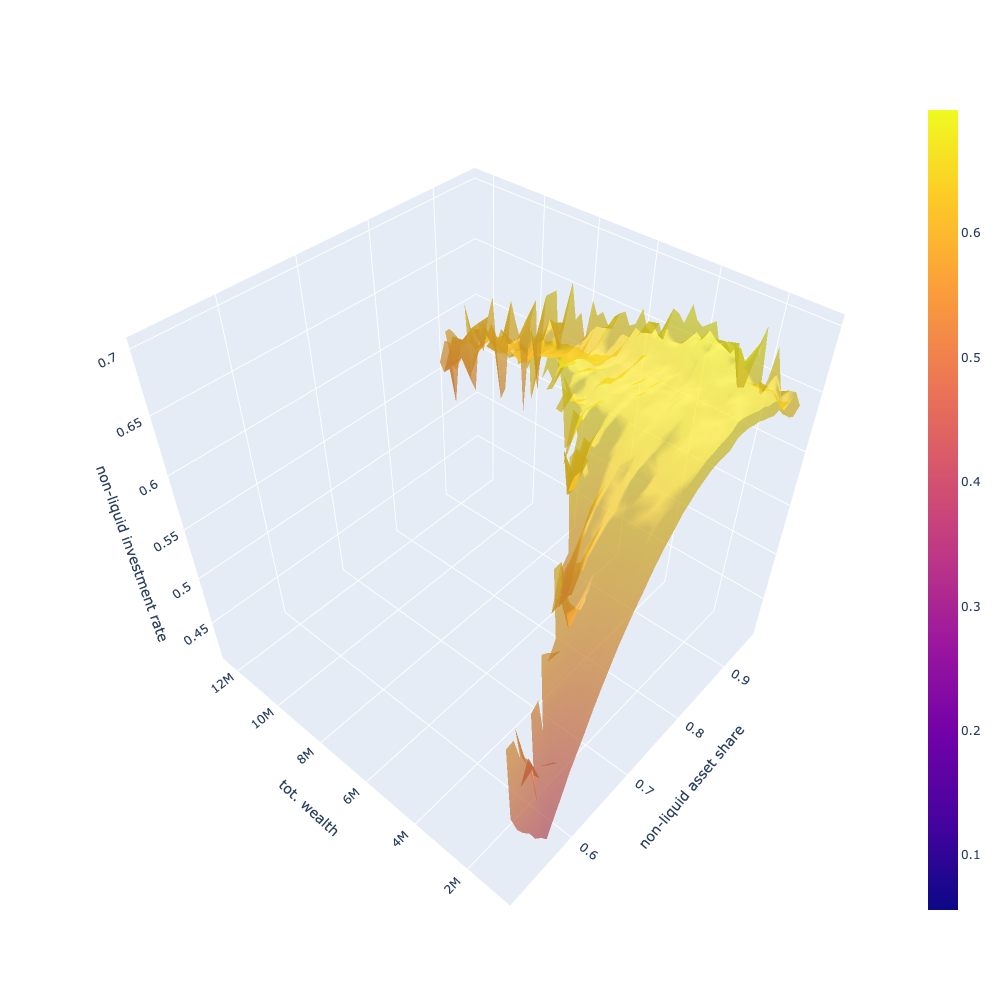}
         \caption{3d Surface Plot of share of non-liquid assets in x-axis, with respect to total asset wealth in y-axis, and corresponding decision of non-liquid asset investment rate in z-axis, the values are smoothed with 9 step moving average for clearer visibility}
         \label{fig:3d_share_wealth_rate}
\end{figure}

The relationship between share of non-liquid asset and age inferred in our model is in par with the literature, furthermore its shape is similar to empirical data presented by Campanale et al. \cite{campanale_life-cycle_2015}. Furthermore, Fig. \ref{fig:3d_share_wealth_rate} reflects a more heterogeneous relationship with higher granularity in comparison to the literature.
Our model provides similar income, consumption, and wealth curves with respect to the literature, and portfolio allocation strategies with higher granularity suitable for heterogeneity and various income processes. A downside of the model is the high level of noise for the portfolio allocation 3d surface \ref{fig:3d_share_wealth_rate}, which might arise from the higher complexity of the model.

\begin{figure}[htp]%
         \centering
         \includegraphics[width=0.5\linewidth]{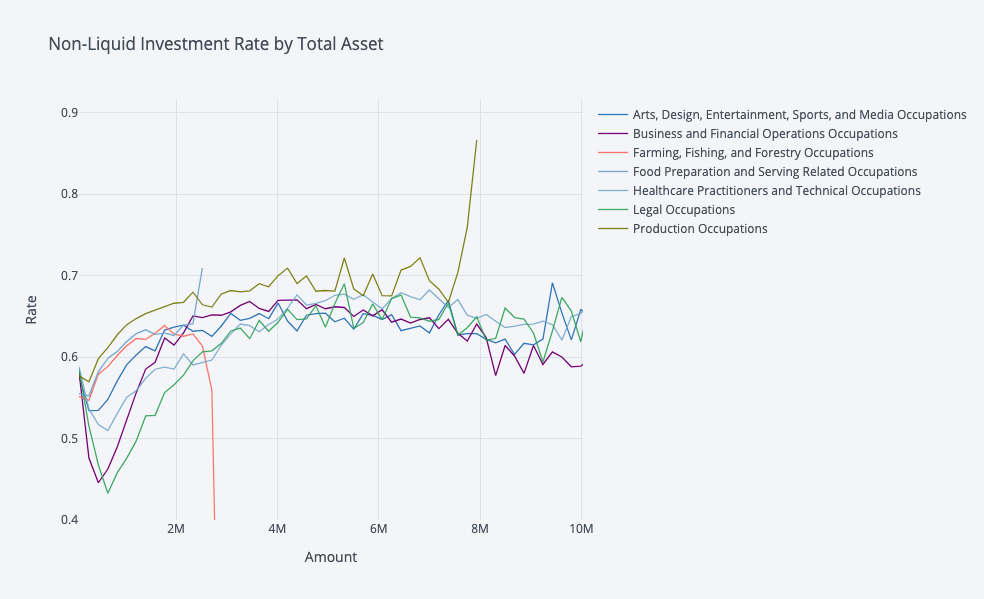}
         \caption{Non liquid investment rate by occupation at amount capped at 10M, the values are smoothed by 30-step moving average and only a selection of occupations are depicted on plot for clear visibility.}
         \label{fig:non_liquid_investment_rate_by_occ_at_asset}
\end{figure}
Contrasting general saving rate and the non-liquid investment rate characteristics of occupations with respect to total assets result in interesting findings.
The non-liquid investment rate by total asset among occupations diverges less than the saving rate by total asset, but still the characteristically differentiating investment strategies are evident in Fig. \ref{fig:non_liquid_investment_rate_by_occ_at_asset}
Saving rate by total asset generally increases for all occupations with more asset, with exponential like increase, then it plateaus and slightly varies with noise. Saving rates by highest total asset amounts fluctuate greatly, which might be due to different dynamics governing their decisions such as capital income or behavioural parameters weighing more themselves rather than income being the determinant of the decisions.

Non-liquid investment rate per total asset is much higher for low income occupations, and only exceptions is for very high amounts of total asset, where higher income occupations can surpass the low income occupations by non-liquid investment rate. This difference is due to a necessity for low-income occupations, which results in lower buffer savings in liquid assets that can be used to finance short-term consumption during periods of unemployment.
This finding can be a guidance for policy makers to come up with instruments to mitigate the risk for income occupations to face higher risk of consumption crisis, such as unemployment benefits, or early career pension contributions from the government to support low-income early career workers.
Non-liquid investment rate seems to have an increasing trend for low-income occupations with respect to total asset. On the other hand, for some of the mid and high occupations, the trend is first a decrease followed by an increase.

\newcommand*{\tabindent}{ \hspace{3mm}}
\begin{table}[htp]
\small
\begin{tabular}{l|llll}
Occupation                                     & Quart 1. & Quart 2. & Quart 3. & Quart 4. \\ \hline
Arts-Design-Entertainment-Sports-and Media &              &              &              &              \\
\tabindent 20-30                           & 0.681        & 0.701        & 0.705        & 0.754        \\
\tabindent 30-40                           & 0.702        & 0.769        & 0.803        & 0.849        \\
\tabindent 40-50                           & 0.696        & 0.745        & 0.805        & 0.863        \\
\tabindent 50-60                           & 0.721        & 0.771        & 0.828        & 0.871        \\
\tabindent 60-70                           & 0.539        & 0.752        & 0.833        & 0.863        \\ \hline
Business and Financial Operations          &              &              &              &              \\
\tabindent 20-30                           & 0.626        & 0.648        & 0.615        & 0.632        \\
\tabindent 30-40                           & 0.653        & 0.671        & 0.712        & 0.774        \\
\tabindent 40-50                           & 0.659        & 0.700        & 0.793        & 0.848        \\
\tabindent 50-60                           & 0.658        & 0.692        & 0.775        & 0.848        \\
\tabindent 60-70                           & 0.547        & 0.695        & 0.759        & 0.841        \\ \hline
Farming-Fishing-and Forestry               &              &              &              &              \\
\tabindent 20-30                           & 0.774        & 0.810        & 0.856        & 0.901        \\
\tabindent 30-40                           & 0.772        & 0.893        & 0.946        & 0.965        \\
\tabindent 40-50                           & 0.758        & 0.854        & 0.938        & 0.965        \\
\tabindent 50-60                           & 0.751        & 0.843        & 0.922        & 0.963        \\
\tabindent 60-70                           & 0.547        & 0.799        & 0.907        & 0.959        \\ \hline
Healthcare Practitioners and Technical     &              &              &              &              \\
\tabindent 20-30                           & 0.644        & 0.655        & 0.653        & 0.680        \\
\tabindent 30-40                           & 0.674        & 0.674        & 0.714        & 0.793        \\
\tabindent 40-50                           & 0.674        & 0.716        & 0.774        & 0.832        \\
\tabindent 50-60                           & 0.692        & 0.738        & 0.793        & 0.851        \\
\tabindent 60-70                           & 0.663        & 0.746        & 0.797        & 0.857        \\ \hline
Legal                                      &              &              &              &              \\
\tabindent 20-30                           & 0.644        & 0.614        & 0.607        & 0.632        \\
\tabindent 30-40                           & 0.659        & 0.649        & 0.717        & 0.802        \\
\tabindent 40-50                           & 0.664        & 0.709        & 0.760        & 0.827        \\
\tabindent 50-60                           & 0.670        & 0.675        & 0.730        & 0.828        \\
\tabindent 60-70                           & 0.630        & 0.699        & 0.791        & 0.859        \\ \hline
Production                                 &              &              &              &              \\
\tabindent 20-30                           & 0.667        & 0.704        & 0.751        & 0.826        \\
\tabindent 30-40                           & 0.682        & 0.770        & 0.843        & 0.900        \\
\tabindent 40-50                           & 0.694        & 0.779        & 0.849        & 0.909        \\
\tabindent 50-60                           & 0.684        & 0.771        & 0.858        & 0.912        \\
\tabindent 60-70                           & 0.589        & 0.757        & 0.839        & 0.904        \\ \hline
All Occupations                            &              &              &              &              \\
\tabindent 20-30                           & 0.680        & 0.723        & 0.743        & 0.735        \\
\tabindent 30-40                           & 0.707        & 0.769        & 0.806        & 0.830        \\
\tabindent 40-50                           & 0.708        & 0.774        & 0.819        & 0.847        \\
\tabindent 50-60                           & 0.703        & 0.774        & 0.817        & 0.856        \\
\tabindent 60-70                           & 0.610        & 0.755        & 0.815        & 0.860     \\ \hline
Campanale et al. TC high             &     &      &        &        \\
\tabindent 20-30                              & 0.077     & 0.471     & 0.467      & 0.577     \\
\tabindent 30-40                              & 0.575    & 0.591     & 0.547      & 0.739     \\
\tabindent 40-50                              & 0.539    & 0.621     & 0.757      & 0.704     \\
\tabindent 50-60                              & 0.70    & 0.765     & 0.791      & 0.698     \\
\tabindent 60-70                              & 0.735    & 0.767     & 0.751       & 0.706     \\
\tabindent 70-80                              & 0.562    & 0.701     & 0.756      & 0.667      \\ \hline
\end{tabular}
   \caption{Occupation and Age vs Share of Non-Liquid Investments for Wealth Quartiles, the results from Campanale et al. with high Transaction Costs(TC) are used for comparison, in our model there are no transfers between non-liquid and liquid assets before retirement so  high transaction cost results are relatively compatible with our model}
    \label{fig:occ_age_vs_invshare_quartile}
\end{table}

Comparing our results with those of Campanale et al. in Table \ref{fig:occ_age_vs_invshare_quartile} with high transaction cost makes it clear that our model refers to a higher non-liquid investment share with respect to total asset portfolio, but there are exceptions where Campanale et al. determine a high non-liquid asset share.
Furthermore, profile specific differences between occupations is evident from varying non liquid asset shares with respect to income quartile and age group.

\subsection{Effects of Behavioural Parameters}
The behavioural parameters of the agents are the consumption utility factor, the shock sensitivity factor, and the individuality factor. These factors capture the behaviour of agents, and they impact how agents perceive, understand, and act in their environment. The consumption utility factor is necessary for quantifying how agents value immediate consumption, which can be interpreted as level of consumerism, or temporal preference and eagerness. The shock sensitivity factor is a parameter helpful for capturing the agent's perception of the consumption change, which can amplify the effects of the changes and force agents to avoid abrupt changes, and an alternative interpretation can be as risk aversion modifier that augments the utility. The individuality factor models the level of influence an agent's social network exerts on the agent. This is achieved by factoring in the information transmitted from the neighbourhood. The increase in the liquid assets reflect a linear increase; on the contrary, the increase of non-liquid assets is exponential due to interest income of the assets. The distribution of outcomes reflect heterogeneous characteristics according to behavioural parameters and the relationship between parameters and outcomes are non-linear. The details can be found in the Appendix.

\clearpage
\section{Conclusion}

We modelled a pension ecosystem, where heterogeneous contributors make consumption and investment decisions with Deep RL, which advances available models by providing better granularity and accounting for profile heterogeneity.

We provide a novel methodology to optimise agent behaviour for consumption and investment between pension savings and liquid cash buffer, which is flexible and can be calibrated to work in various scenarios and capture agent heterogeneity. Our model does not need an explicit formulation of the income process and can work with empirical data. 

Our research represents a first step of end-to-end modelling of pension ecosystem, to provide a highly capable model to optimise contributor and other actor behaviour, in the existence of high variability and heterogeneity, as well as dynamic environment. We successfully devised optimal contributor portfolio allocation strategies between nonliquid pension savings and liquid cash buffer, as well as optimal consumption decision, which can be calibrated with behavioural parameters of agents. We accomplish this by minimising the consumption crisis periods of agents and maximising the retirement savings.

\section{Future Work}
Future work can be planned in four dimensions. One dimension is the representativeness and information richness, which can be enhanced by more in depth shock embedding for capturing how a shock affects a specific agent, considering shocks such as financial crashes, epidemics, and natural catastrophes such as earthquakes. 
A second dimension is a more in depth representation of each profession using vector representation of professions, which will capture the job similarity and to some extent the relationship between professions. Utilising a more comprehensive graph encompassing real geographical neighbourhoods, peer network, or employment network, as well as synthethic networks with well-known structures such as small world will provide more extensive capabilities to model the world.
The thirds dimension of future work is introducing more actors in the multi-agent model including companies as employers and business-to-business relationships, which will provide a model less leaning on statistical calibration via real-world data, and capture second-order effects and endogenous properties. Another pivotal agent that can increase the impact and self-consistency of the simulated environment is the government acting as a tax regulator.
The fourth dimension is technical improvement to the simulation system to use vectorised parallel architectures such as GPUs, which will enable us to test more extensive market conditions more efficiently and in shorter time. 

\clearpage
\bibliographystyle{unsrt}
\bibliography{citations}
\clearpage
\section{Appendix}
\subsection{Model Card}
\begin{table}[htp]
\begin{tabular}{|ll|}
\hline
\multicolumn{2}{|l|}{Simulation Parameters}                                                                \\ \hline
\multicolumn{1}{|l|}{Parallel Environment Count}                               & 32                                       \\ \hline
\multicolumn{1}{|l|}{Income Calibration Data}           & USA CPS 2019 Median weekly earnings      \\ \hline
\multicolumn{1}{|l|}{Unemployment Duration Data}& USA CPS 2019 Unemployment duration table \\ \hline
\multicolumn{1}{|l|}{use\_min\_max\_scaler}                     & 1                                        \\ \hline
\multicolumn{1}{|l|}{time steps}                               & 1000                                     \\ \hline
\multicolumn{1}{|l|}{consumption\_crisis\_penalty}              & 100000                                   \\ \hline
\multicolumn{1}{|l|}{invalid\_action\_penalty\_modifier}        & 1000                                     \\ \hline
\multicolumn{1}{|l|}{retirement\_age}                           & 65                                       \\ \hline
\multicolumn{1}{|l|}{retirement\_salary\_multiplier}            & 0.8                                      \\ \hline
\multicolumn{1}{|l|}{death\_rate}                         & USA SSA Actuarial Life Table                   \\ \hline
\multicolumn{1}{|l|}{Agent States}                              & \begin{tabular}[c]{@{}l@{}}"OCC\_CODE" , "age", "income",\\ "consumption\_utility\_factor",\\ "shock\_sensitivity\_factor","individuality\_factor",\\"non\_liquid\_asset", "liquid\_asset"{]}\end{tabular} \\ \hline
\multicolumn{2}{|l|}{Market Parameters}                                                                    \\ \hline
\multicolumn{1}{|l|}{monthly\_market\_interest\_rate}           & 0                                        \\ \hline
\multicolumn{1}{|l|}{CPI}                                       & 0                                        \\ \hline
\multicolumn{1}{|l|}{monthly\_non\_liquid\_asset\_return\_rate} & 0.0125                                   \\ \hline
\multicolumn{1}{|l|}{monthly\_liquid\_asset\_return\_rate}      & 0.0025                                   \\ \hline
\multicolumn{1}{|l|}{monthly\_minimum\_consumption}             & 1073 (2021 USA Poverty Guidelines)                                    \\ \hline
\multicolumn{1}{|l|}{monthly\_minimum\_wage}                    & 1160                                     \\ \hline
\multicolumn{2}{|l|}{ML Parameters}                                                                        \\ \hline
\multicolumn{1}{|l|}{batch\_size}                               & 14656                                       \\ \hline
\multicolumn{1}{|l|}{n\_lstm}                                   & 128                                      \\ \hline
\multicolumn{1}{|l|}{n\_steps}                                  & 1                                        \\ \hline
\multicolumn{1}{|l|}{}                                          &                                          \\ \hline
\end{tabular}
\caption{Parameters}
\label{tab:parameter-table}
\end{table}
\clearpage
\subsection{Nomenclature}
\nomenclature{$c_{i,t}$}{Accumulated consumption value}
\nomenclature{$\mathbb{C}_{i,t}$}{Instant Consumption}
\nomenclature{$\mathbb{E}_{i,t}$}{Current Earnings}
\nomenclature{$m_{i,t}$}{Minimum consumption}
\nomenclature{$p_{i,t}$}{Accumulated penalty}
\nomenclature{$P_{i,t}$}{Instant Penalty for Consumption Crisis}
\nomenclature{$\eta$}{Degree of non-linearity of crra function}
\nomenclature{$q_{i,t}$}{Consumption Utility Factor}
\nomenclature{$\kappa_{i,t}$}{Shock Sensitivity Factor}
\nomenclature{$A_{a,b}$}{1 if there is an edge a-$>$ b}
\nomenclature{$\delta(x,y)$}{Kronecker Delta}
\nomenclature{$\psi$}{Consumption Crisis Penalty Modifier}
\nomenclature{$\zeta$}{Invalid Action Penalty Modifier}
\printnomenclature
\subsection{Neural Architecture}
\begin{figure}[htp]%
         \centering
         \includegraphics[width=0.89\linewidth]{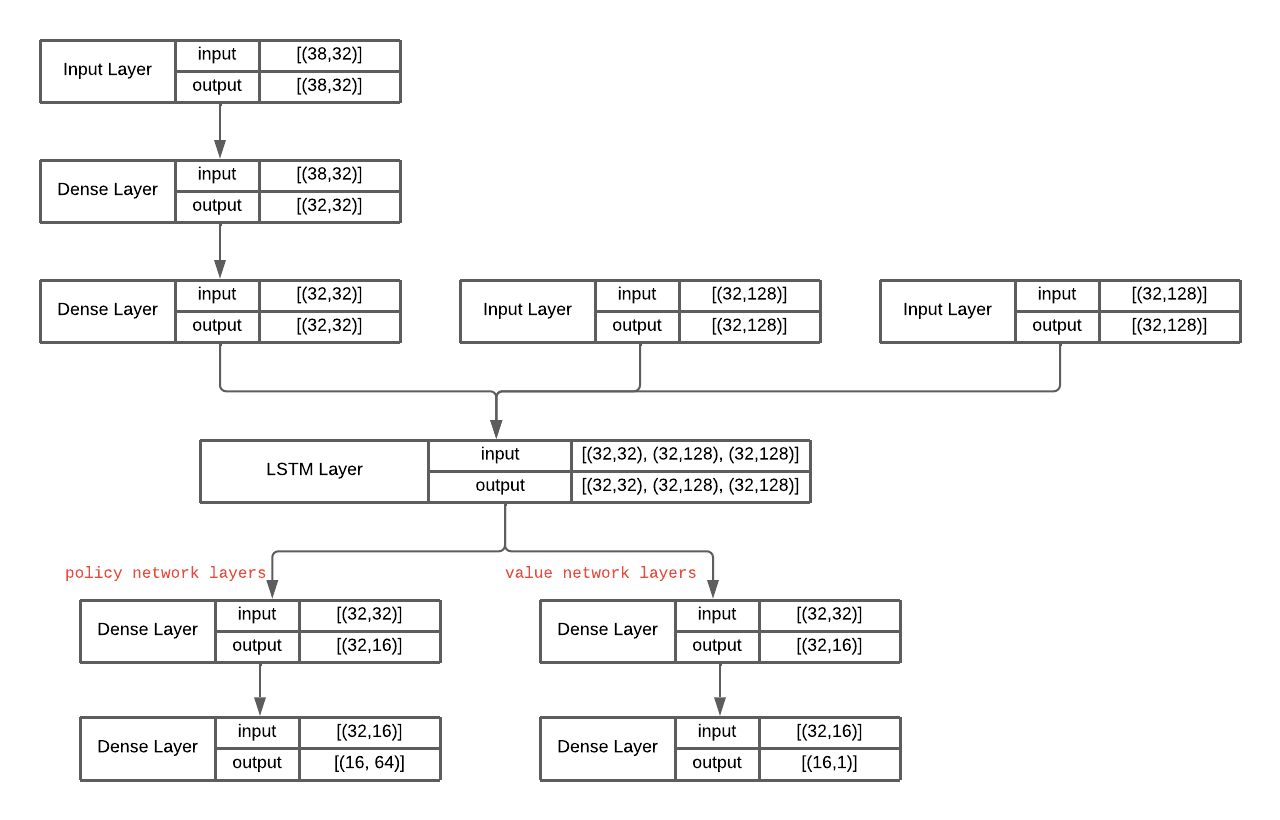}
         \caption{Neural Architecture}
         \label{fig:ml_plot}
\end{figure}
\clearpage
\subsection{Graph Plot}
\begin{figure}[htp]%
         \centering
      \includegraphics[width=1\linewidth]{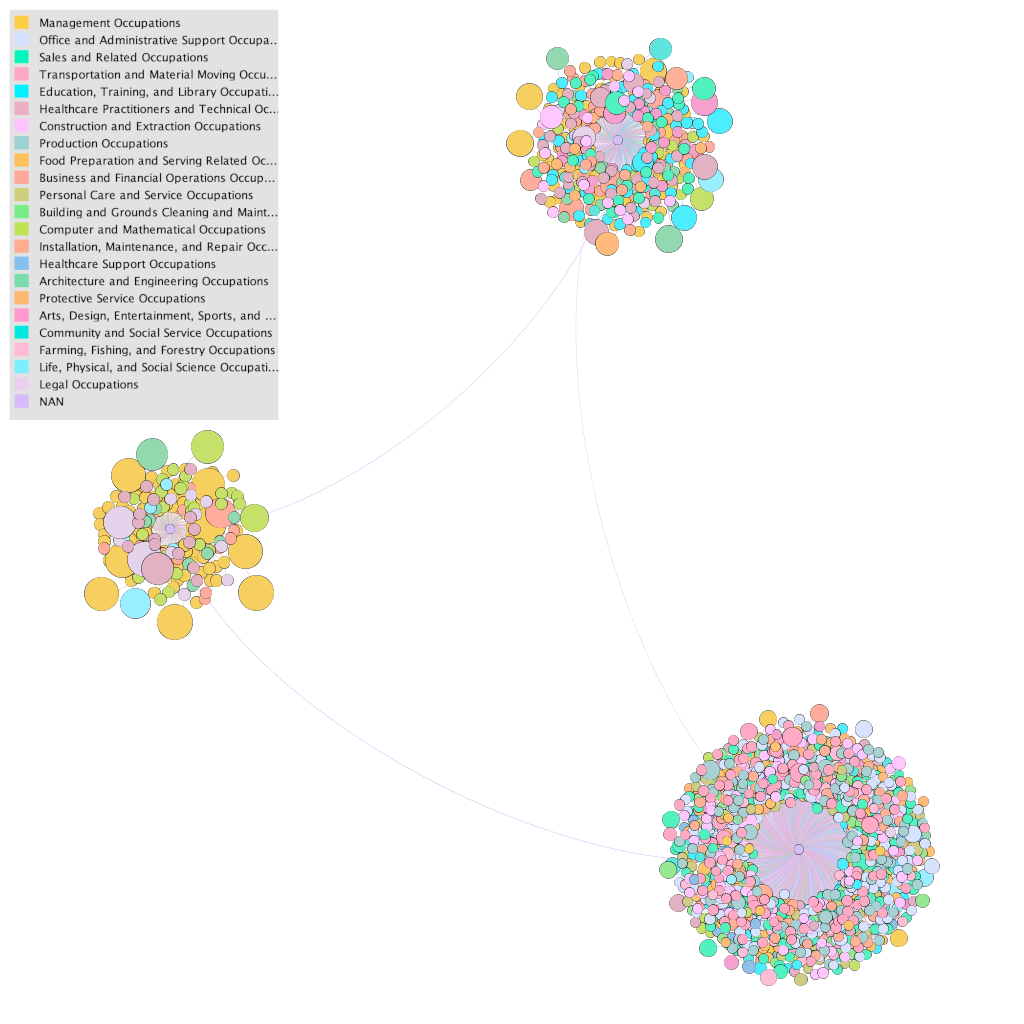}
         \caption{Graph Plot where occupations are reflected with colours and income is reflected with the size of nodes, the graph consists of three sub-groups representing three neighbourhood with three differing income levels as high mid and low income, central nodes of each group are representing the neighbourhoods.}
         \label{fig:graph_plot}
\end{figure}
\clearpage

\subsection{Cross-sectional analysis}

\begin{figure}[htp]%
         \centering
         \subfloat[Liquid Investment Rate]{\includegraphics[width=0.5\linewidth]{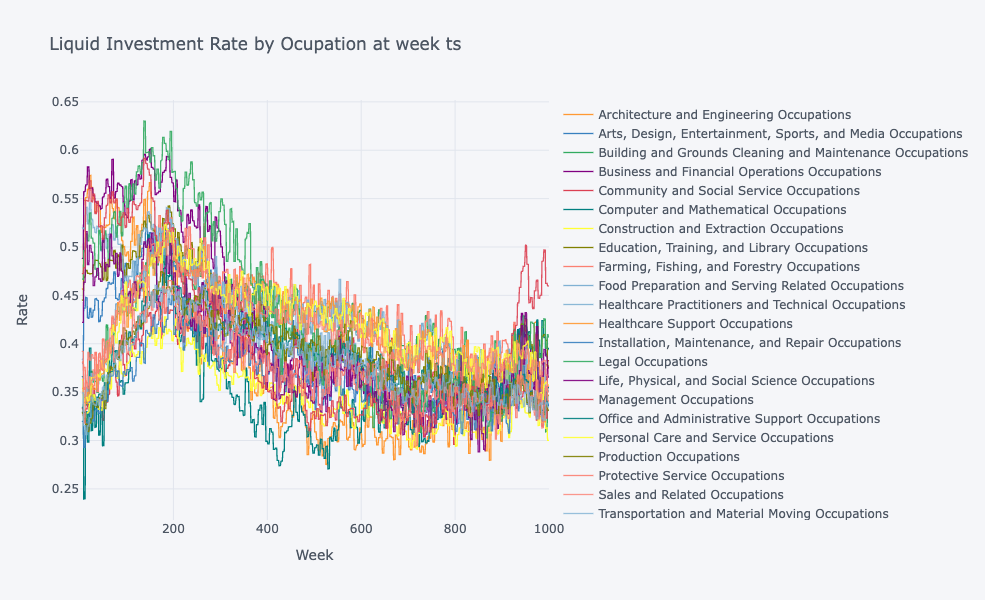}}
         \subfloat[Non-Liquid Investment Rate]{\includegraphics[width=0.5\linewidth]{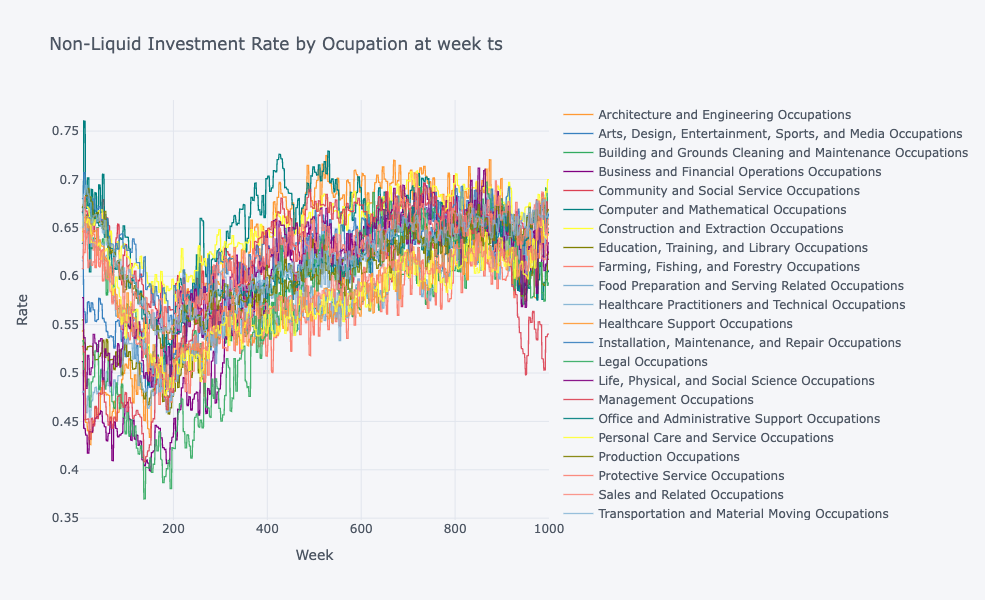}}
         \caption{Liquid and Nonliquid Investment rate by Occupation at week ts}
         \label{fig:liquid_nonliquid_investment_rate_by_ocupation_at_week_ts}
\end{figure}
\begin{figure}[htp]%
         \centering
         \subfloat[Liquid Investment Rate]{\includegraphics[width=0.5\linewidth]{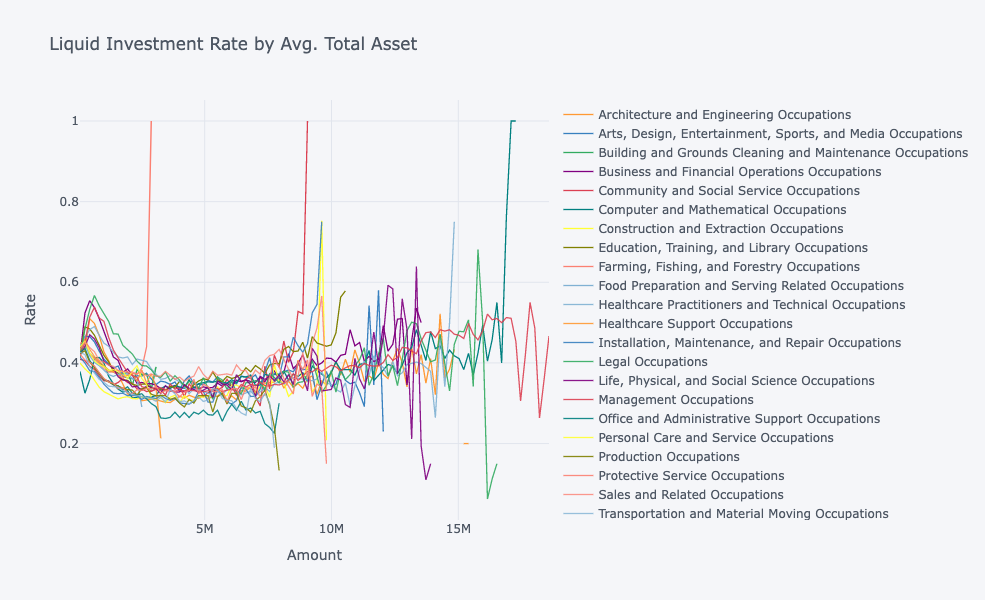}}
         \subfloat[Non-Liquid Investment Rate]{\includegraphics[width=0.5\linewidth]{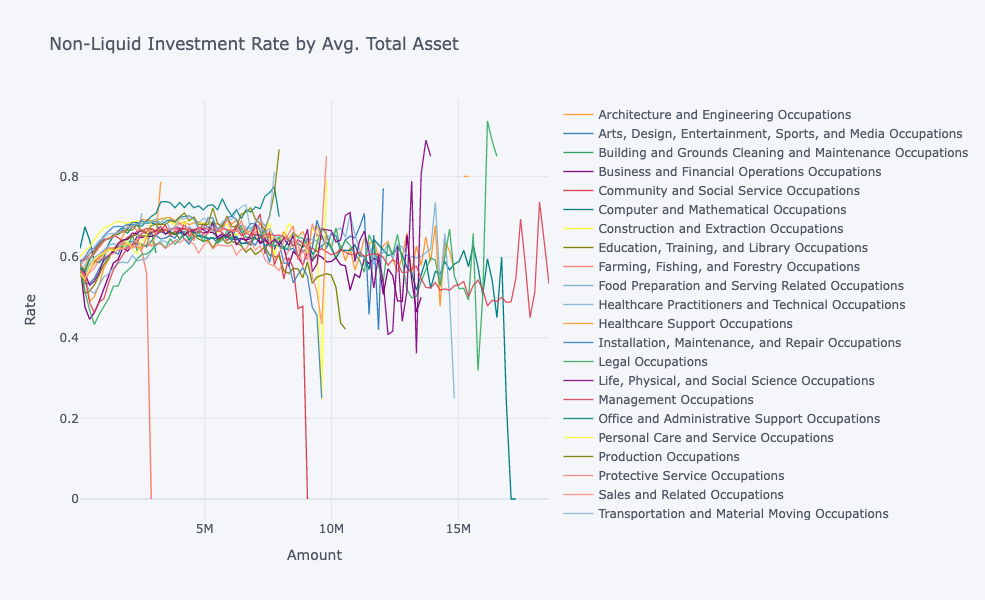}}
         \caption{liquid and nonliquid investment rate by occ at amount}
         \label{fig:liquid_nonliquid_investment_rate_by_ocupation_at_amount}
\end{figure}

The Fig. \ref{fig:liquid_nonliquid_investment_rate_by_ocupation_at_amount} reflects the relationship between liquid investment rate and wealth for amounts less than 5M USD. 3 distinctive behaviours are observable, one is Computer and Mathematical Occupations, which start at lowest liquid investment rate, the other group represents the majority of the occupations representing most of low- and mid-income occupations, which start at nearly 35\% but then lower their liquid investment rates when the total wealth increases, the third group consist mostly of the high income occupations such as Management Occupations which increase their liquid investment rate with total asset increase until nearly 500K USD, at that point they start to decrease their liquid investment rate.

\begin{figure}[htp]%
         \centering
         \subfloat[Liquid Asset]{\includegraphics[width=0.5\linewidth]{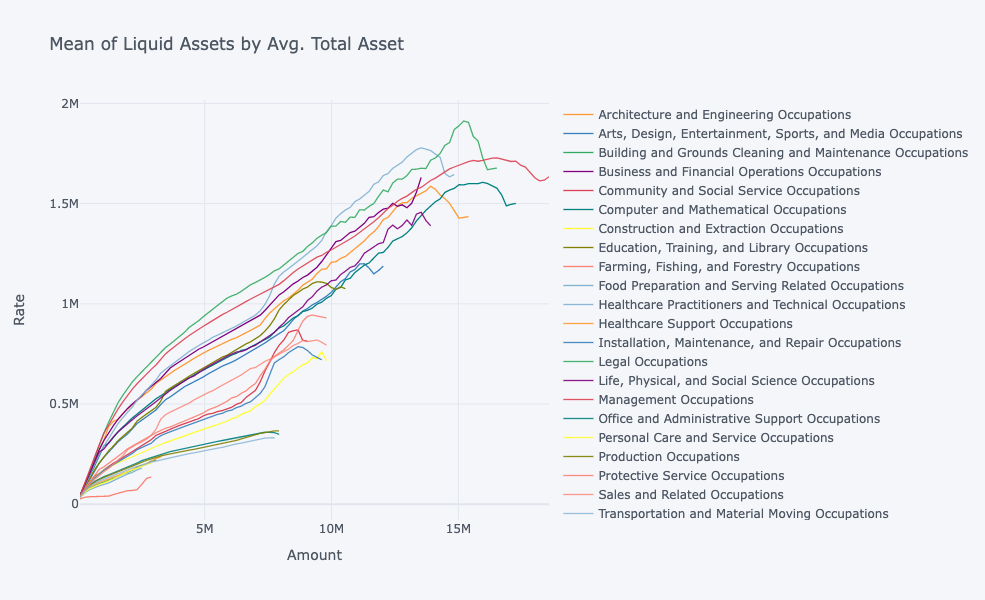}}
         \subfloat[Non-Liquid Asset]{\includegraphics[width=0.5\linewidth]{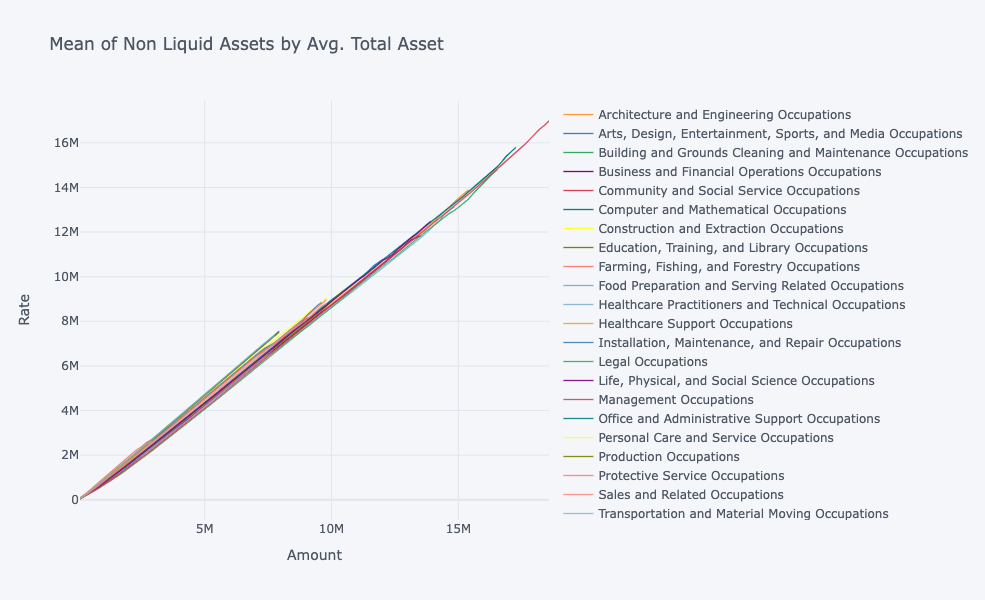}}
         \caption{Liquid and Nonliquid Assets by occupation at Total Amount}
         \label{fig:liquid_nonliquid_assets_by_ocupation_at_amount}
\end{figure}

Fig. \ref{fig:liquid_nonliquid_assets_by_ocupation_at_amount} shows that the increase in liquid assets slows with increasing total wealth, which reflects the fact that the need for security buffer savings decrease and reward of illiquid asset is higher. On the contrary, the increase of no liquid assets with respect to the total wealth increase speeds up at higher amounts and converges to a stable linear trajectory.

The distribution of assets with respect to age on Fig. \ref{fig:liquid_nonliquid_assets_by_ocupation_at_age} highly differentiates according to the occupation, where Management and Legal Occupations possess highest value of assets and Farming, Fisheries and Food Preparation Occupations possess lowest level of assets and also reflect lowest change of assets with age, which points out that asset differentiation with respect to age depends heavily on the occupation type, where some occupations show greatly changing income asset values and some occupations can provide minimal savings opportunity, due to income merely sufficing to finance current consumption during workforce participation.

\begin{figure}[htp]%
         \centering
         \subfloat[Liquid Asset]{\includegraphics[width=0.5\linewidth]{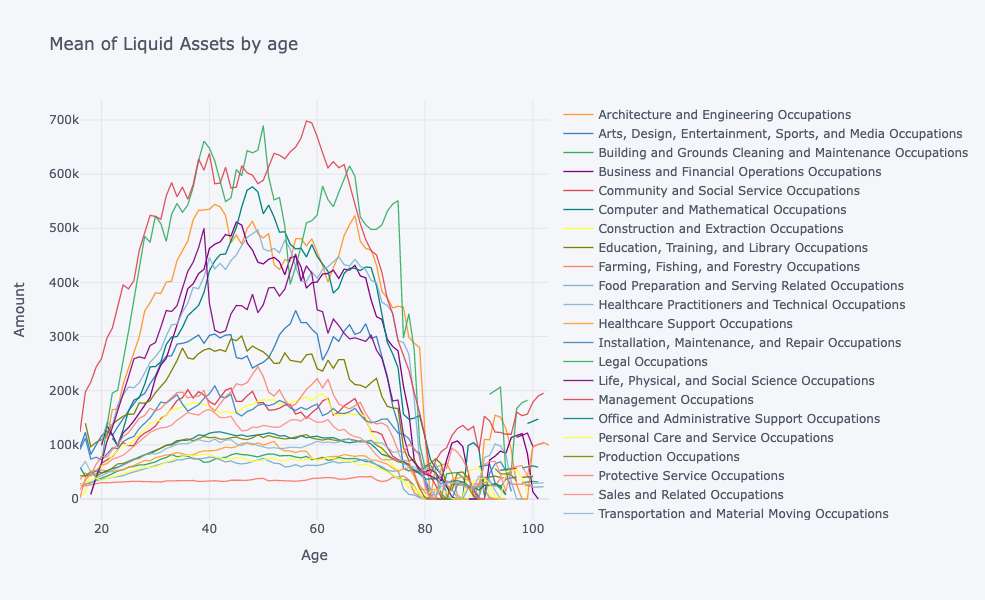}}
         \subfloat[Non-Liquid Asset]{\includegraphics[width=0.5\linewidth]{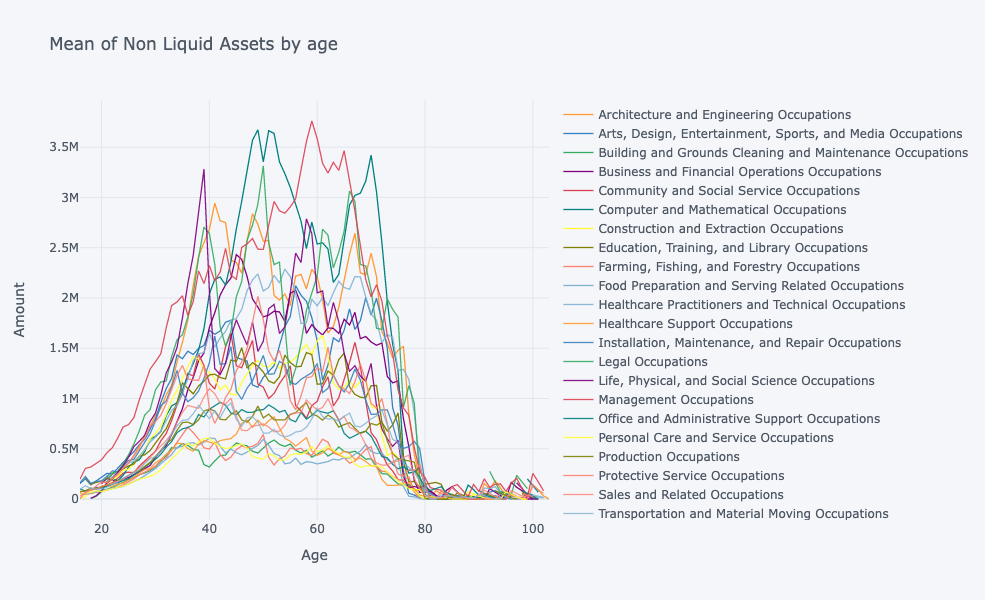}}
         \caption{Liquid and Nonliquid assets by occupation at age}
         \label{fig:liquid_nonliquid_assets_by_ocupation_at_age}
\end{figure}
\begin{figure}[htp]%
         \centering
         \subfloat[Liquid Asset]{\includegraphics[width=0.33\linewidth]{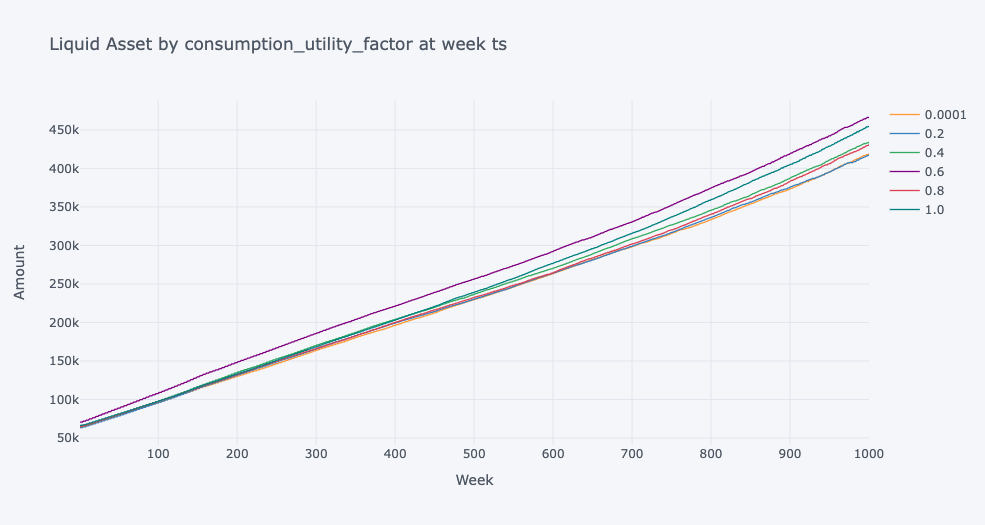}}
         \subfloat[Non-Liquid Asset]{\includegraphics[width=0.33\linewidth]{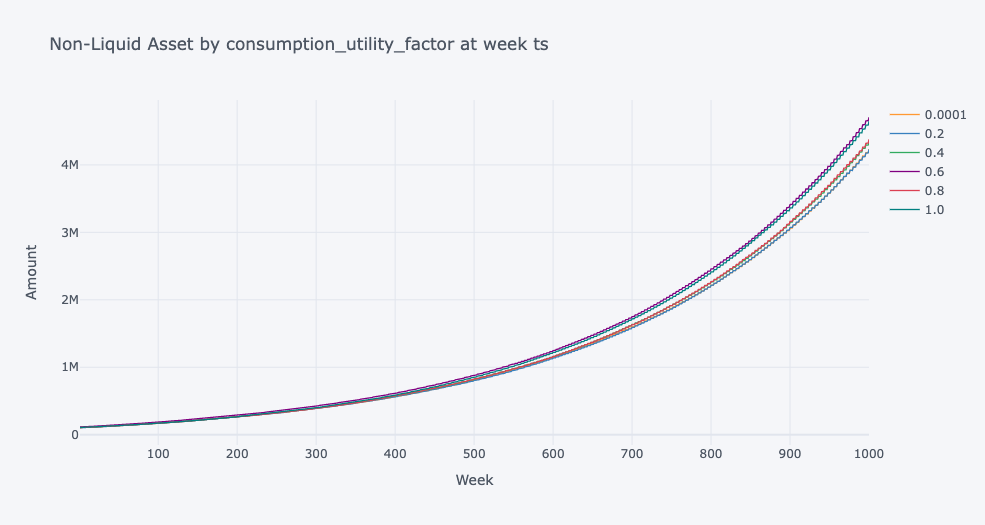}}
          \subfloat[Total Asset]{\includegraphics[width=0.33\linewidth]{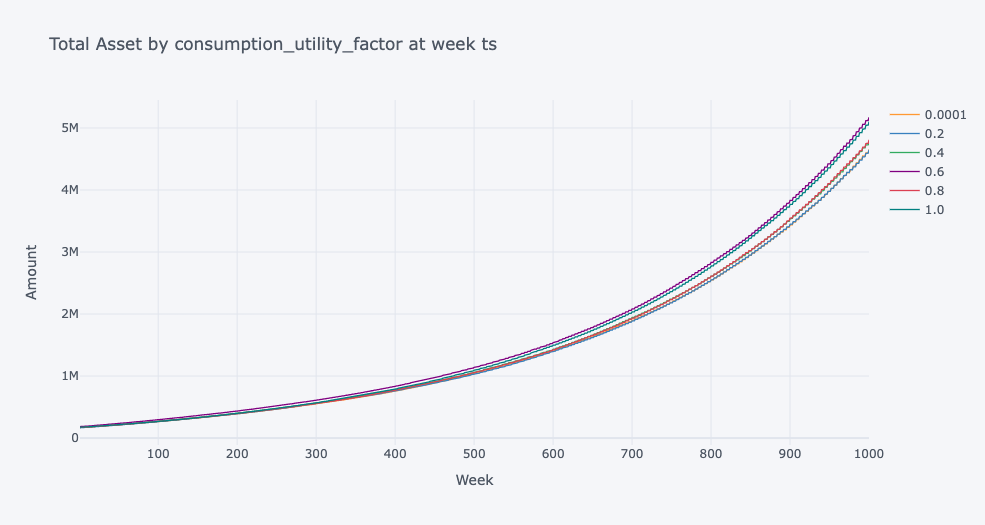}}
         \caption{Total Asset, Nonliquid Asset, Liquid Asset at week ts by Consumption Utility factor}
         \label{fig:liquid_nonliquid_total_assets_by_consumption_utility_factor_at_week}
\end{figure}
\clearpage
\subsection{Tables}

\begin{table}[htp]
\resizebox{\textwidth}{!}{%
\begin{tabular}{l|lll}
occupation title                               & saving rate & \begin{tabular}[c]{@{}l@{}}non liquid\\ investment rate\end{tabular} & \begin{tabular}[c]{@{}l@{}}share of non\\ liquid investments\end{tabular} \\ \hline
Architecture and Engineering                   & 0.578       & 0.607                      & 0.730                           \\
Arts, Design, Entertainment, Sports, and Media & 0.422       & 0.587                      & 0.756                           \\
Building and Grounds Cleaning and Maintenance  & 0.207       & 0.576                      & 0.777                           \\
Business and Financial Operations              & 0.512       & 0.557                      & 0.714                           \\
Community and Social Service                   & 0.363       & 0.623                      & 0.773                           \\
Computer and Mathematical                      & 0.596       & 0.643                      & 0.769                           \\
Construction and Extraction                    & 0.374       & 0.649                      & 0.787                           \\
Education, Training, and Library               & 0.404       & 0.574                      & 0.743                           \\
Farming, Fishing, and Forestry                 & 0.160       & 0.569                      & 0.845                           \\
Food Preparation and Serving Related           & 0.218       & 0.574                      & 0.790                           \\
Healthcare Practitioners and Technical         & 0.533       & 0.576                      & 0.737                           \\
Healthcare Support                             & 0.210       & 0.576                      & 0.777                           \\
Installation, Maintenance, and Repair          & 0.381       & 0.635                      & 0.786                           \\
Legal                                          & 0.519       & 0.550                      & 0.703                           \\
Life, Physical, and Social Science             & 0.471       & 0.603                      & 0.732                           \\
Management                                     & 0.590       & 0.584                      & 0.722                           \\
Office and Administrative Support              & 0.274       & 0.608                      & 0.778                           \\
Personal Care and Service                      & 0.208       & 0.573                      & 0.778                           \\
Production                                     & 0.267       & 0.608                      & 0.782                           \\
Protective Service                             & 0.396       & 0.622                      & 0.777                           \\
Sales and Related                              & 0.282       & 0.578                      & 0.767                           \\
Transportation and Material Moving             & 0.269       & 0.613                      & 0.789                           \\ \hline
\end{tabular}
}
    \caption{Occupation vs Rates, the saving rate denotes to the average monthly saving rate of the members of each occupation, the non liquid investment rate denotes the average of the decided rate of allocating monthly savings to non liquid investments for each occupation, the share of non liquid investments denotes the share of non liquid assets with respect to all of the investments averaged for each occupation}
    \label{fig:occ_vs_rates}
\end{table}

\begin{table}[]
\resizebox{\textwidth}{!}{%
\begin{tabular}{l|l|l|l|l}
age & non liquid investment rate & share of non liquid investments & consumption rate & Coco et al. consumption rate \\ \hline
21  & 0.551                      & 0.677                           & 0.648            & 0.884                       \\
22  & 0.546                      & 0.684                           & 0.634            & 0.915                       \\
23  & 0.553                      & 0.690                           & 0.629            & 0.948                       \\
24  & 0.554                      & 0.701                           & 0.628            & 0.976                       \\
25  & 0.556                      & 0.710                           & 0.623            & 0.996                       \\
26  & 0.558                      & 0.717                           & 0.611            & 0.998                       \\
27  & 0.562                      & 0.728                           & 0.608            & 0.999                       \\
28  & 0.567                      & 0.736                           & 0.607            & 0.999                       \\
29  & 0.571                      & 0.743                           & 0.609            & 0.998                       \\
30  & 0.572                      & 0.750                           & 0.606            & 0.996                       \\
31  & 0.579                      & 0.756                           & 0.598            & 0.987                       \\
32  & 0.584                      & 0.764                           & 0.595            & 0.979                       \\
33  & 0.587                      & 0.771                           & 0.596            & 0.972                       \\
34  & 0.590                      & 0.777                           & 0.591            & 0.966                       \\
35  & 0.592                      & 0.781                           & 0.591            & 0.962                       \\
36  & 0.594                      & 0.782                           & 0.588            & 0.960                       \\
37  & 0.596                      & 0.783                           & 0.583            & 0.959                       \\
38  & 0.597                      & 0.786                           & 0.578            & 0.959                       \\
39  & 0.597                      & 0.786                           & 0.575            & 0.960                       \\
40  & 0.600                      & 0.789                           & 0.571            & 0.962                       \\
41  & 0.597                      & 0.787                           & 0.563            & 0.963                       \\
42  & 0.601                      & 0.785                           & 0.562            & 0.965                       \\
43  & 0.603                      & 0.785                           & 0.561            & 0.966                       \\
44  & 0.603                      & 0.787                           & 0.562            & 0.966                       \\
45  & 0.604                      & 0.786                           & 0.559            & 0.966                       \\
46  & 0.605                      & 0.788                           & 0.561            & 0.966                       \\
47  & 0.604                      & 0.787                           & 0.559            & 0.965                       \\
48  & 0.605                      & 0.789                           & 0.557            & 0.964                       \\
49  & 0.603                      & 0.788                           & 0.555            & 0.963                       \\
50  & 0.605                      & 0.788                           & 0.555            & 0.963                       \\
51  & 0.609                      & 0.789                           & 0.557            & 0.964                       \\
52  & 0.607                      & 0.787                           & 0.558            & 0.966                       \\
53  & 0.607                      & 0.785                           & 0.564            & 0.970                       \\
54  & 0.605                      & 0.785                           & 0.563            & 0.976                       \\
55  & 0.606                      & 0.786                           & 0.568            & 0.985                       \\
56  & 0.604                      & 0.787                           & 0.569            & 0.996                       \\
57  & 0.604                      & 0.787                           & 0.569            & 1.011                       \\
58  & 0.604                      & 0.790                           & 0.573            & 1.029                       \\
59  & 0.606                      & 0.790                           & 0.572            & 1.051                       \\
60  & 0.606                      & 0.789                           & 0.578            & 1.077                       \\
61  & 0.609                      & 0.788                           & 0.578            & 1.107                       \\
62  & 0.606                      & 0.788                           & 0.572            & 1.142                       \\
63  & 0.607                      & 0.784                           & 0.572            & 1.180                       \\
64  & 0.606                      & 0.785                           & 0.570            & 1.223                       \\ \hline
\end{tabular}%
}
    \caption{Age vs Rates: Consumption rates are defined as consumption amount divided to income. Consumption rates are compared to literature by extracting values from plots of Coco et al. \cite{cocco_consumption_2005}, their research differs by our work such that the income values exclude contributions toward pension income, and savings are used as a mean to finance consumption deficit especially during retirement, so during retirement there are positive consumption rates, which mean that the pension deficit is financed by spending savings. This definition difference causes consumption rates to be much higher.}
    \label{fig:age_vs_rates}
\end{table}

\begin{table}[htp]
\begin{tabular}{l|lllll}
Occupation                                     & 20-30 & 30-40 & 40-50 & 50-60 & 60-70 \\ \hline
Architecture and Engineering                   & 0.669    & 0.68     & 0.658    & 0.668    & 0.325    \\
Arts, Design, Entertainment, Sports, and Media & 0.465    & 0.448    & 0.445    & 0.484    & 0.236    \\
Building and Grounds Cleaning and Maintenance  & 0.237    & 0.237    & 0.236    & 0.235    & 0.118    \\
Business and Financial Operations              & 0.556    & 0.602    & 0.594    & 0.599    & 0.281    \\
Community and Social Service                   & 0.447    & 0.422    & 0.433    & 0.41     & 0.2      \\
Computer and Mathematical                      & 0.639    & 0.672    & 0.671    & 0.674    & 0.319    \\
Construction and Extraction                    & 0.419    & 0.435    & 0.432    & 0.42     & 0.188    \\
Education, Training, and Library               & 0.456    & 0.471    & 0.472    & 0.436    & 0.211    \\
Farming, Fishing, and Forestry                 & 0.181    & 0.18     & 0.176    & 0.184    & 0.09     \\
Food Preparation and Serving Related           & 0.236    & 0.235    & 0.235    & 0.236    & 0.114    \\
Healthcare Practitioners and Technical         & 0.598    & 0.615    & 0.591    & 0.578    & 0.308    \\
Healthcare Support                             & 0.236    & 0.235    & 0.234    & 0.234    & 0.12     \\
Installation, Maintenance, and Repair          & 0.446    & 0.427    & 0.434    & 0.412    & 0.212    \\
Legal                                          & 0.603    & 0.584    & 0.592    & 0.573    & 0.322    \\
Life, Physical, and Social Science             & 0.56     & 0.549    & 0.556    & 0.535    & 0.286    \\
Management                                     & 0.683    & 0.692    & 0.692    & 0.68     & 0.346    \\
Office and Administrative Support              & 0.31     & 0.315    & 0.317    & 0.311    & 0.152    \\
Personal Care and Service                      & 0.238    & 0.236    & 0.237    & 0.236    & 0.116    \\
Production                                     & 0.309    & 0.305    & 0.31     & 0.303    & 0.147    \\
Protective Service                             & 0.429    & 0.437    & 0.445    & 0.426    & 0.212    \\
Sales and Related                              & 0.334    & 0.313    & 0.323    & 0.312    & 0.157    \\
Transportation and Material Moving             & 0.307    & 0.302    & 0.315    & 0.298    & 0.152 \\ \hline
\end{tabular}
 \caption{Saving Rate by Occupation and Age}
    \label{fig:saving_rate_by_occ_age}
\end{table}

\clearpage
\subsection{Raw Plots and Behavioural Parameter Effect Plot}
\begin{figure}[htp]%
         \centering
         \subfloat[Mean Income]{\includegraphics[width=0.5\linewidth]{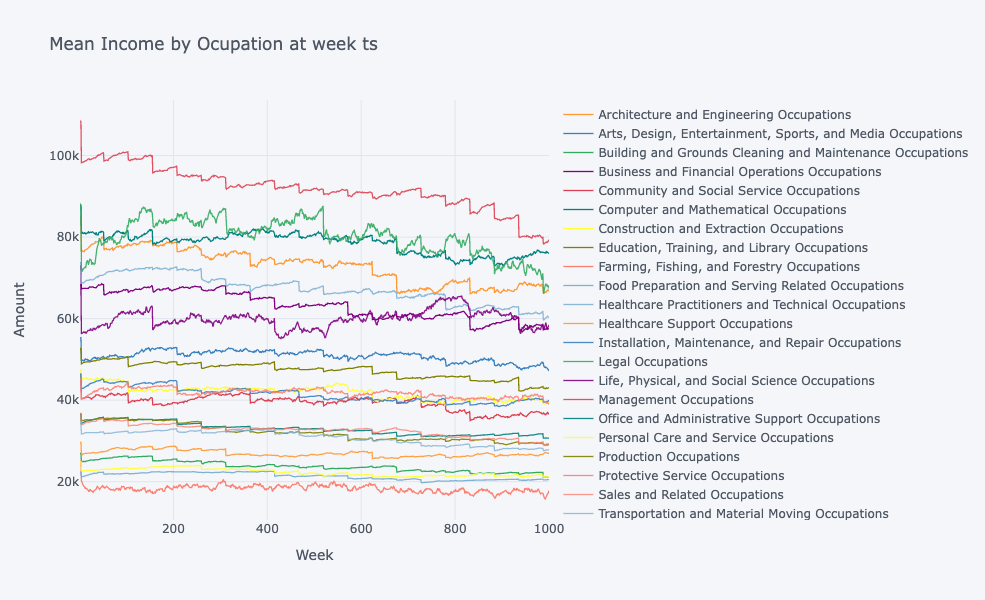}}
         \subfloat[Unemployment]{\includegraphics[width=0.5\linewidth]{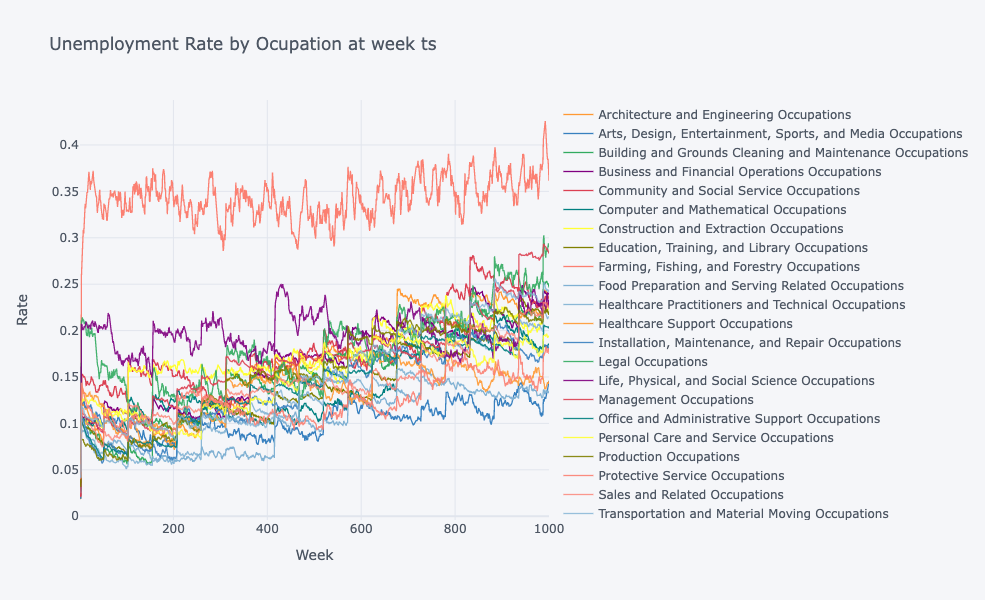}}
         \caption{Mean Income and Unemployment by Occupation at week ts}
         \label{fig:income_unemployment_by_occ_at_week_all}
\end{figure}
\begin{figure}[htp]%
         \centering
         \subfloat[Week]{\includegraphics[width=0.5\linewidth]{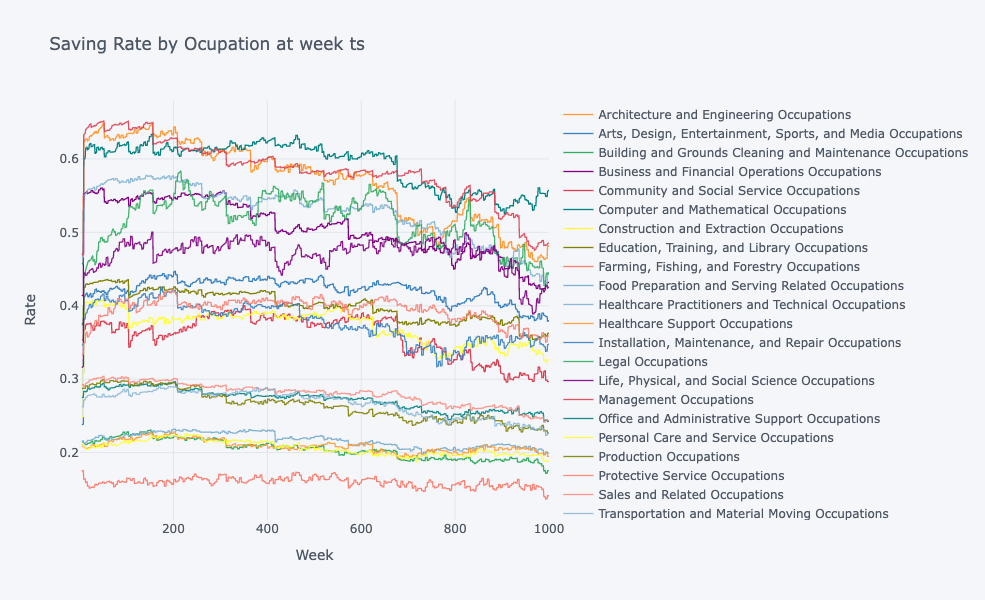}}
         \subfloat[Total Asset Amount]{\includegraphics[width=0.5\linewidth]{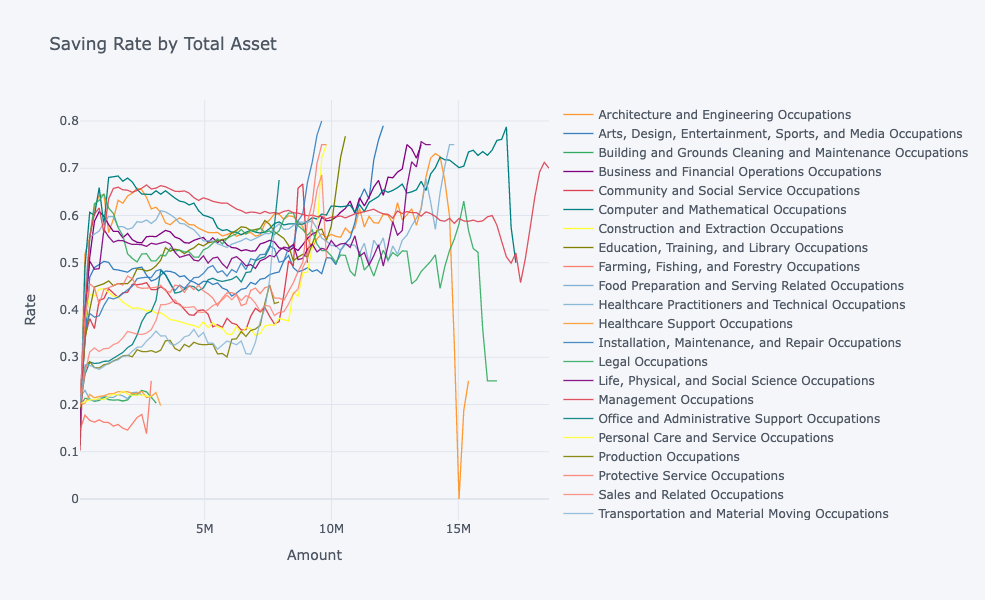}}
         \caption{Saving Rate by Occupation at week ts and Saving Rate by Occupation at amount capped at 10M}
         \label{fig:saving_rate_by_occ_at_week_asset_all}
\end{figure}
\begin{figure}[htp]%
         \centering
      \includegraphics[width=0.55\linewidth]{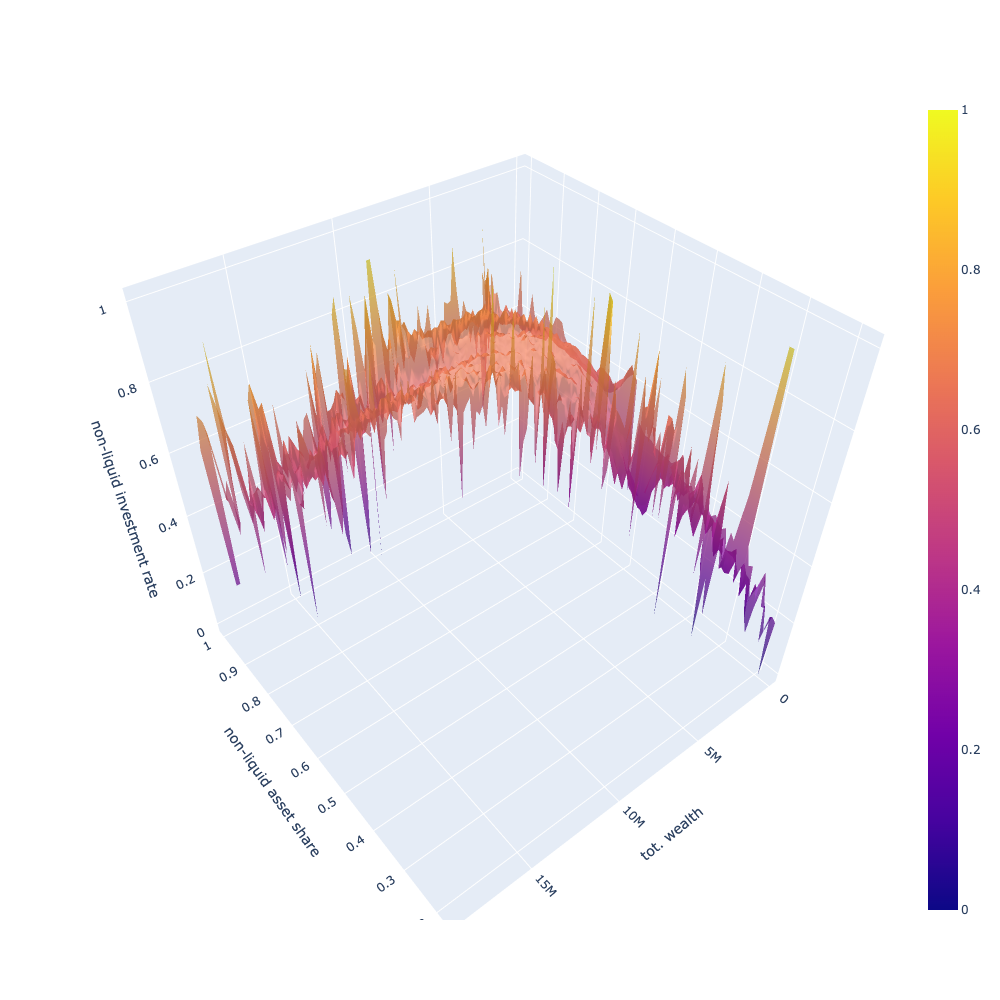}
         \caption{3d Surface Plot of share of non-liquid assets in x-axis, with respect to total asset wealth in y-axis, and corresponding decision of non-liquid asset investment rate in z-axis}
         \label{fig:3d_share_wealth_rate_raw}
\end{figure}

\begin{figure}[htp]%
         \centering
      \includegraphics[width=0.85\linewidth]{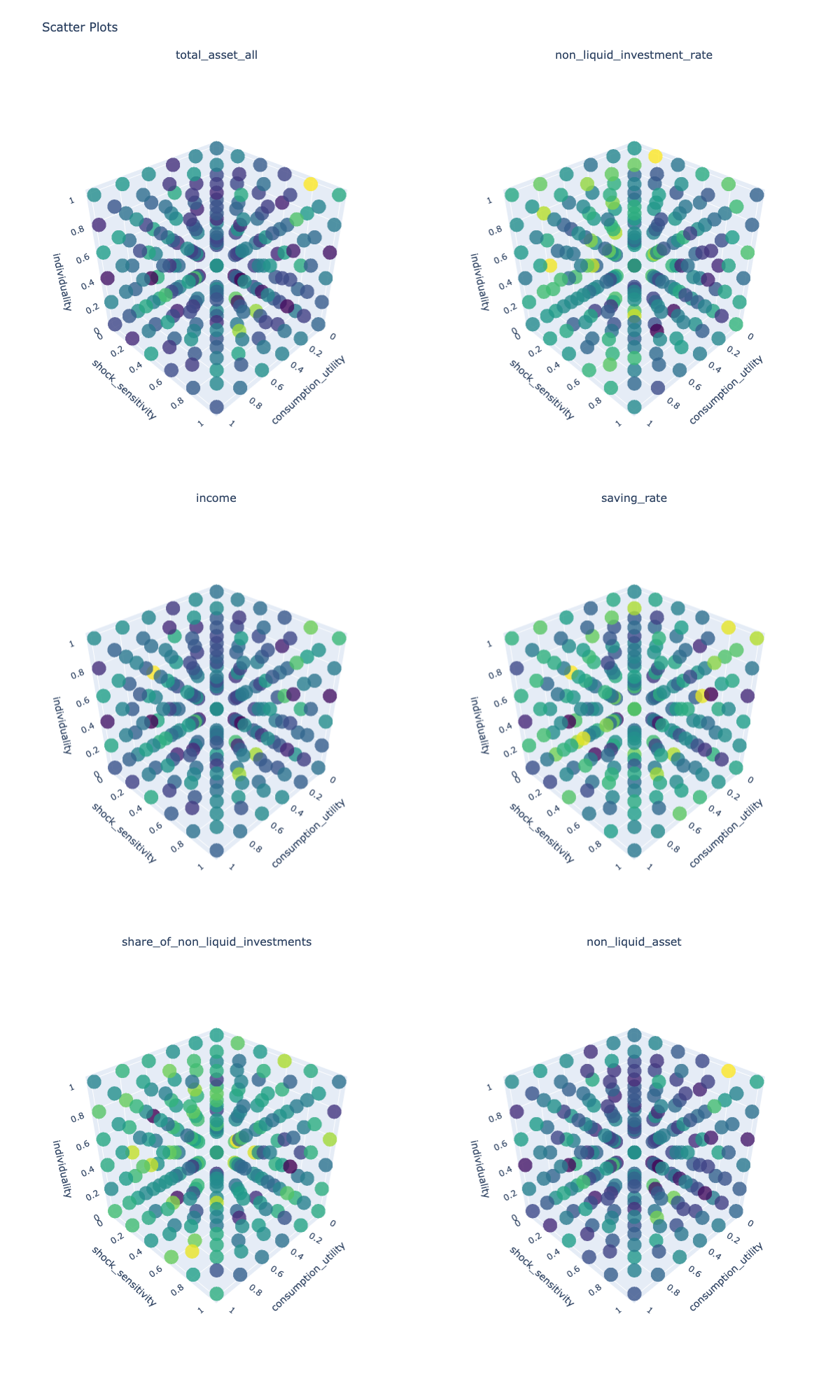}
         \caption{3D Scatter Plot of each indicator relative to the behavioural parameters of the agents, where dark blue indicate lower values and light yellow indicate higher values, which reflect how the parameters are effecting the values such as accumulated assets, investment rates or share of non liquid assets. The income vs parameters plot is provided for convenience, the income itself is not effected by the behavioural parameters}
         \label{fig:3d_parameter_plot}
\end{figure}

\end{document}